\documentclass[VANCOUVER,STIX1COL,doublespace]{WileyNJD-v2}

\newcommand\BibTeX{{\rmfamily B\kern-.05em \textsc{i\kern-.025em b}\kern-.08em
T\kern-.1667em\lower.7ex\hbox{E}\kern-.125emX}}

\articletype{Article Type}%

\received{<day> <Month>, <year>}
\revised{<day> <Month>, <year>}
\accepted{<day> <Month>, <year>}

\begin{document}

\title{Parallel Performance of Algebraic Multigrid Domain Decomposition (AMG-DD)}

\author[1]{Wayne B. Mitchell*}

\author[1]{Robert Strzodka}

\author[2]{Robert D. Falgout}

\authormark{Wayne Mitchell \textsc{et al}}

\address[1]{\orgdiv{ZITI}, \orgname{Universit\"at Heidelberg}, \orgaddress{\state{Baden-W\"urttemberg}, \country{Germany}}}

\address[2]{\orgdiv{CASC}, \orgname{Lawrence Livermore National Lab}, \orgaddress{\state{California}, \country{USA}}}

\corres{*Wayne Mitchell, ZITI, Im Neuenheimer Feld 368, 5. OG, 69120 Heidelberg. \email{wayne.b.mitchell@gmail.com}}

\presentaddress{Present address}

\abstract[Abstract]{Algebraic multigrid (AMG) is a widely used scalable solver and preconditioner for large-scale linear systems resulting from the discretization of a wide class of elliptic PDEs. While AMG has optimal computational complexity, the cost of communication has become a significant bottleneck that limits its scalability as processor counts continue to grow on modern machines. This paper examines the design, implementation, and parallel performance of a novel algorithm, Algebraic Multigrid Domain Decomposition (AMG-DD), designed specifically to limit communication. The goal of AMG-DD is to provide a low-communication alternative to standard AMG V-cycles by trading some additional computational overhead for a significant reduction in communication cost. Numerical results show that AMG-DD achieves superior accuracy per communication cost compared to AMG, and speedup over AMG is demonstrated on a large GPU cluster.}


\jnlcitation{\cname{%
\author{Mitchell W.B.}, 
\author{R. Strzodka}, and 
\author{R.D. Falgout}} (\cyear{2020}), 
\ctitle{Parallel Performance of Algebraic Multigrid Domain Decomposition (AMG-DD)}, \cjournal{}, \cvol{}.}

\maketitle


\section{Introduction}

Algebraic multigrid (AMG) \cite{Brandt:1985um, Stuben:1986vg, McCormick:2016vp} is a widely used scalable solver and preconditioner for large-scale linear systems resulting from the discretization of a wide class of elliptic PDEs. AMG scales well to very large problem sizes \cite{Baker:2012ko} due to its optimal computational complexity: application of an AMG cycle has $O(N)$ computational cost, where $N$ is the size of the linear system to be solved. The cost of communication, however, has become a significant bottleneck that limits the scalability of AMG as processor counts continue to grow on modern machines \cite{Gahvari:2011dh}. In fact, it is well documented that for large processor counts, an AMG V-cycle may spend more time on coarse levels of the hierarchy than the finest level. This difficulty is due to increased communication costs on the coarse levels even though there is exponentially less computational work there compared to the finest level \cite{Bienz:2016ii}. This result is in direct conflict with one of the most general guiding principles of multigrid methods: that the coarse grids should be inexpensive. Thus, in order to maintain the efficiency of AMG on modern supercomputers with ever increasing parallelism, work must be done to ameliorate its communication costs. This paper continues development of algebraic multigrid domain decomposition (AMG-DD) \cite{Anonymous:KA9t5Fed}, a novel algorithm built on top of AMG and designed with the express purpose of achieving similar or better convergence with significantly reduced communication at the cost of some redundant storage and computation.

There are several approaches available when considering the reduction of communication costs in AMG that may be divided into three broad categories that may each be employed in conjunction with one another: optimization and algorithm development for parallel implementation; modification of the AMG algorithm itself through various changes to the AMG hierarchy; and use of entirely different cycling structures or algorithms based on the AMG hierarchy such as AMG-DD. 

The first category includes changes to the parallel implementation that avoid any changes to the underlying AMG algorithm itself and thus exactly preserve convergence and other algorithmic properties. In \cite{Anonymous:ycC62NK_}, for example, the authors develop an assumed partition algorithm for efficiently determining interprocessor communication patterns when global partitioning information is unavailable. The applicability of this algorithm is quite general, but, in particular, it may be used to speed up the AMG setup phase where communication patterns for coarse-grid matrices must be established. In \cite{Bienz:uc}, the authors develop node-aware communication patterns for general sparse matrix-vector multiplication that may be applied to speed up the solve phase of AMG. These approaches to communication reduction are the least invasive in that there is no change to the AMG algorithm itself, so speedup may be achieved with no fear of degraded convergence. 

A more invasive category of approaches involves changing the AMG hierarchy itself. The cost of communication is generally most problematic on coarse grids in the middle of the AMG hierarchy. On the finest grids, communication patterns are typically regular and fairly sparse (if the problem comes from, for example, the discretization of a PDE that is then decomposed among processors), and the amount of computational work is typically very large compared to the required communication. On the coarsest grids, on the other hand, the total number of degrees of freedom can be significantly less than the number of processors, meaning that many processors are inactive and the total amount of communication is small. That leaves the mid-range of the hierarchy, where increasing complexity of the coarse-grid operators leads to large communication stencils and, thus, large communication costs that dominate the relatively small computational cost on these levels. On systems where communication is a dominant cost, AMG may actually spend \emph{more} time on these mid-range coarse levels than the fine grid (see, for example, numerical results from \cite{Bienz:2016ii}), contradicting a basic heuristic of multigrid methods that the coarse grids should be cheap relative to the fine grid. Thus, efforts to ameliorate communication cost through modification of the AMG hierarchy have focused on sparsification of the coarse-grid operators. Modified coarsening and interpolation schemes are developed in \cite{DeSterck:2008fc, DeSterck:2006et} that are designed to produces sparser coarse-grid operators with correspondingly sparser communication patterns. In \cite{Yang:2010ii}, aggressive coarsening is used in conjunction with improved long-distance interpolation in order to both sparsify the coarse grids and reduce the total number of grid levels. More direct approaches to sparsification through elimination of matrix entries resulting in non-Galerkin coarse-grid operators are explored in \cite{Falgout:2014fw, Bienz:2016ii}. In each of these cases, the sparsified hierarchies yield faster AMG V-cycles at the cost of a degraded V-cycle convergence factor. Thus, more (but faster) V-cycles may be required to solve a given problem to a desired tolerance. This tradeoff means that the benefits of using these modified AMG hierarchies may be very problem dependent: significant speedup is observed for some problems, whereas convergence is destroyed for others.

Finally, there has been exploration of altogether different algorithms based on AMG (or multigrid methods in general) that aim to replace typical multigrid cycling. 
A main underlying principle here is the attempt to add parallelism and/or reduce communication by decoupling pieces of the multigrid cycle. 
Additive multigrid methods, for example, add extra concurrency compared to traditional multiplicative methods by decoupling the levels of the multigrid hierarchy. 
Traditional additive multigrid methods suffer from worse convergence compared to their multiplicative counterparts, though more advanced variants on additive multigrid that recover good convergence exist, such as the asynchronous fast adaptive composite grid methods described in \cite{Lee:2004cj} or the additive multigrid methods using modified smoothed interpolation operators described in \cite{Vassilevski:2014eh}.
Algebraic multigrid domain decomposition (AMG-DD) retains a multiplicative multigrid cycling approach, but it decouples the spatial decomposition of the problem among processors, employing independent cycling on each processor to reduce communication by storing fully overlapping composite grid structures. Similar algorithms employing overlapping composite grids have been explored in the context of geometric multilevel methods and adaptive mesh refinement. The parallel meshing algorithm from \cite{Bank:1999uq} uses the idea of storing additional overlapping information on each processor in order to determine adaptively refined meshes with reduced communication but then relies on subsequent conventional domain decomposition or multigrid iterations to perform the solve. In \cite{Mitchell:2004hz,Mitchell:2016vg}, a multigrid solver based on full domain partitions and a hierarchical basis is developed that requires communication only twice per cycle. The range-decomposition algorithm developed in \cite{Appelhans:2017,Mitchell:2018fl} enables fully independent nested-iteration solves with adaptive mesh refinement on each processor. In contrast to these algorithms, AMG-DD operates in a purely algebraic context, providing a linear solver that requires only the fine-grid matrix equation with no need for additional discretization information or infrastructure.

AMG-DD was originally presented in \cite{Anonymous:KA9t5Fed}, in which the authors provide some basic motivating theory, performance models, and test results using a model serial code for small problem sizes. This paper continues the development of AMG-DD through further modification and design of the algorithm, implementation of a scalable parallel code, and empirical study of algorithm behavior and parallel performance. The crucial algorithmic development presented here is an efficient algebraic fast adaptive composite (AlgFAC) cycle with minimal storage and communication requirements. In addition, the communication algorithm proposed in \cite{Anonymous:KA9t5Fed} is implemented and modified to avoid the redundant communication present in the original algorithm. Section \ref{amg_sec} gives a brief review of AMG and is followed by a description of AMG-DD in Section \ref{amgdd_sec}. Section \ref{fac_sec} explains the newly developed AlgFAC cycle used as the primary computational routine of AMG-DD, and Section \ref{res_comm_sec} gives further detail on the primary communication routine of AMG-DD. Finally, Section \ref{results_sec} studies optimal parameter choices for AMG-DD, examines the scaling behavior of AMG-DD, and demonstrates speedup of AMG-DD over AMG on a large GPU cluster.


\section{Review of AMG}
\label{amg_sec}

Algebraic multigrid (AMG) \cite{Brandt:1985um, Stuben:1986vg, McCormick:2016vp} generates a multigrid hierarchy and then solves a given linear system, $Au=f$, using only the information given in the matrix $A$. The given linear system comprises the finest level of the hierarchy, level $l=0$, so denote $A = A_0$, $u = u_0$, and $f = f_0$, and let $N = N_0$ be the number of degrees of freedom in this system. Generating a set of coarse-grid degrees of freedom may be done either by splitting the fine-grid degrees of freedom into disjoint sets of fine points ($F$-points) and coarse points that are repeated on the coarse grid ($C$-points), as done in "classical" AMG, or by grouping several degrees of freedom and treating them as a single point on the coarse grid, as done in "smoothed aggregation" AMG \cite{Vanek:1996dca}. This paper treats only classical AMG, although nothing prevents AMG-DD from being applied to smoothed aggregation. 
In classical AMG, the splitting of the degrees of freedom into $C$- and $F$-points is typically accomplished via a coloring algorithm that seeks to generate a maximal subset of $C$-points such that $C$-points do not strongly influence each other but do strongly influence the remaining $F$-points, where "strong influence" is determined by the relative size of entries in $A_0$. Given such a splitting with $N_1$ $C$-points, an $N_0\times N_1$ interpolation operator, $P_0$, is constructed based on connections in the matrix, $A_0$. Typically, an interpolated value at a $C$-point is simply injected from the coarse grid, while a value at an $F$-point is obtained by applying interpolation weights (determined based on connections in $A_0$) to values at $C$-points in the $F$-point's interpolatory set. Classically, the interpolatory set contains $C$-points within distance 1 through the graph of $A_0$, but interpolation operators based on longer-distance interpolatory sets are also now commonly used \cite{DeSterck:2008fc,Yang:2010ii}. An $N_1\times N_0$ restriction operator, $R_0$, may also be generated separately, but often is set as $R_0 = P_0^T$. Once the interpolation and restriction operators are defined, the coarse-grid operator is then formed by the triple matrix product, $A_1 = R_0A_0P_0$. This process may then be repeated recursively on level $l = 1, 2,...,L$, splitting the level $l$ degrees of freedom into $C$- and $F$-points, defining $P_l$ and $R_l$, and then forming $A_{l+1} = R_lA_lP_l$, until a sufficiently coarse level, $L$, on which $A_L$ is small enough to be easily handled by relaxation or a direct method such as Gaussian elimination. 

Construction of the interpolation, restriction, and coarse-grid operators comprises the setup phase of AMG. The solve phase then involves cycling on the hierarchy of grids by performing some relaxation on a grid level (usually some variation on an easily applied, pointwise iterative method like Jacobi or Gauss-Seidel), then either restricting a residual equation or interpolating a correction to move up or down through the hierarchy. Algorithm \ref{vcyclealg} shows pseudocode and lists communications for an AMG V$(\nu_1, \nu_2)$-cycle, the most commonly used cycle structure, which is simply a downsweep performing $\nu_1$ relaxations on each level followed by an upsweep performing $\nu_2$ relaxations on each level. For simple relaxation methods like Jacobi and Gauss-Seidel, the cost of relaxation on level $l$ is equivalent to performing a matrix-vector multiplication (mat-vec) with $A_l$. Thus, the cost associated with an AMG V$(\nu_1,\nu_2)$-cycle is the cost of $(1+\nu_1+\nu_2)$ mat-vecs with $A_l$ ($\nu_1 + \nu_2$ relaxations and a residual calculation) plus a mat-vec with $P_l$ for interpolation and a mat-vec with $R_l$ for restriction on each level, $l$. When running an AMG cycle on a parallel system, degrees of freedom on each level are partitioned among processors. In order to perform a mat-vec, each processor must then communicate with nearest processor neighbors to exchange "halo" data, i.e. vector data at points that are distance 1 (through the graph of the matrix) away from the owned, on-processor points. As stated in the introduction, on the finest levels of the hierarchy, this communication cost is usually dominated by computational cost, but growing operator complexities on coarser levels can lead to a sharp increase in the number of neighbors each processor must communicate with in order to perform a mat-vec. Thus, these coarse-level communication costs may dominate the total cost of the entire cycle.

\begin{algorithm}[H]
\caption{AMG V$(\nu_1,\nu_2)$-cycle}
\label{vcyclealg}
\begin{algorithmic}
\begin{multicols}{2}
   \For {Downsweep: $l = 0,...,(L-1)$}
      \State{Relax $\nu_1$ times on $A_lu_l = f_l$.}
      \State{Restrict residual: $f_{l+1} \leftarrow R_l(f_l - A_lu_l)$}.
      \State{Initialize: $u_{l+1} \leftarrow \mathbf{0}$}.
   \EndFor
   \State{Coarse solve: $u_L \leftarrow A_L^{-1}f_L$}.
   \For {Upsweep: $l = (L-1),...,0$}
      \State{Interpolate correction: $u_l \leftarrow u_l + P_lu_{l+1}$}.
      \State{Relax $\nu_2$ times on $A_lu_l = f_l$.}
   \EndFor
   \columnbreak
   \\
   \\
   (mat-vec communications with $A_l$) \\
   (mat-vec communications with $A_l$ and $R_l$) \\
   (no communication) \\
   \\
   (mat-vec communication with $A_L$) \\
   \\
   (mat-vec communications with $P_l$) \\
   (mat-vec communications with $A_l$)
\end{multicols}
\end{algorithmic}
\end{algorithm}


\section{AMG-DD algorithm description}
\label{amgdd_sec}

\begin{figure}[t!]
\centering
\includegraphics[width=0.3\textwidth]{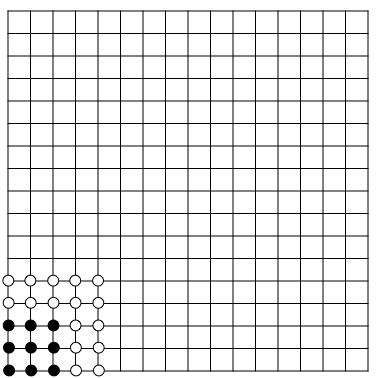}
\hspace{0.2 cm}
\includegraphics[width=0.3\textwidth]{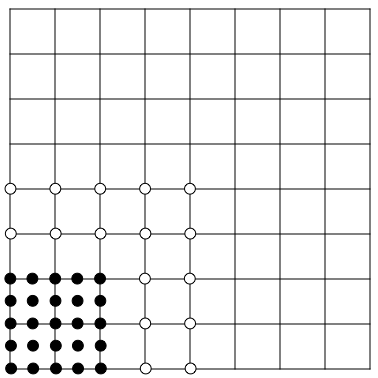}
\hspace{0.2 cm}
\includegraphics[width=0.3\textwidth]{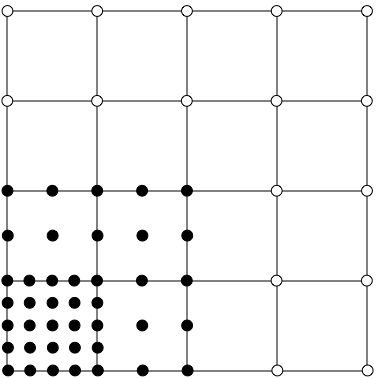}

\caption{Illustration of composite grid construction on a regular, 2D grid with $\eta_l = 2$ on all levels.}
\label{compGridCreation}
\end{figure}

Algebraic multigrid domain decomposition (AMG-DD) seeks to reduce the communication cost of the AMG solve phase by constructing global composite grids on each processor based on the underlying AMG hierarchy and that processor's owned partition of the degrees of freedom, referred to here as that processor's subdomain. These composite grids are generated by moving through the graphs of the operators, $A_l$, on each level of the AMG hierarchy, as illustrated in Figure \ref{compGridCreation}. First, choose some padding value for each level, denoted $\eta_l$. Then, beginning on the finest level, a processor's composite grid includes its subdomain (the black points in the left pane of Figure \ref{compGridCreation}) plus additional points within distance $\eta_0$ through the graph of $A_0$ (the white points in the left pane of Figure \ref{compGridCreation}). All $C$-points from the resulting set are then included in the composite grid on level 1 plus all points within distance $\eta_1$ through the graph of $A_1$ (the white points in the middle pane of Figure \ref{compGridCreation}). This process of coarsening and then extending through the graph of $A_l$ by distance $\eta_l$ is then repeated for all of the coarse grids. The coarsest-level padding, $\eta_L$, should be chosen large enough that each processor stores the entire coarsest grid (depending on the paddings chosen and the size of the coarsest grids, this may be true for a few of the coarsest grids). Thus, each processor stores a composite grid that has finest-level information over its subdomain plus additional information at composite-grid points that extend further away over the global domain on coarser levels. Processors may independently perform cycling on these composite grids through an algebraic fast adaptive composite (AlgFAC) cycling routine (described in detail in Section \ref{fac_sec}).

Algorithm  \ref{amgdd:alg} shows pseudocode and lists communications for an AMG-DD cycle. The cycle provides a correction to the residual equation, $A_0 \delta u_0 = f_0 - A_0u_0$, for some current fine-grid iterate, $u_0$. The first step is to calculate a current residual and restrict that residual to all levels of the hierarchy (using regular, distributed mat-vecs). Then a more complicated algorithm (described in detail in Section \ref{res_comm_sec}) is used to communicate residuals such that each processor obtains current residuals at all of their composite grid points (note that before this step, processors only have access to current residuals in their subdomain). These residuals are then used as the right-hand sides for a local solve via AlgFAC cycles on the composite grids in which each processor, $p$, obtains its own correction term, $\delta u_0^p$. The correction to the global solution is applied locally on each processor, that is, each processor updates $u_0 \leftarrow u_0 + \delta u_0^p$ over its fine-grid subdomain, $\Omega_0^p$, using fine-grid values from its locally calculated composite-grid solution. Since the composite grids stored on each processor cover the global domain at some coarse level, the local solves performed on each processor may be thought of as solving a composite representation of the global problem. This composite representation is (hopefully) accurate to the global problem in the processor's subdomain, where it has finest-level information, and so the global correction produced by AMG-DD is patched together from these processor subdomains.

\begin{algorithm}[H]
\caption{AMG-DD cycle}
\label{amgdd:alg}
\begin{algorithmic}
\State{}
\vspace{-0.5 cm}
\begin{multicols}{2}
   \State{Calculate fine-grid residual: $r_0 \leftarrow f_0-A_0u_0$.}
   \State{Restrict residual to all levels: $r_{l+1} \leftarrow R_lr_l, \forall l$.}
   \State{Obtain residual values at all composite-grid points.}
   \State{Perform local AlgFAC cycles to obtain $\delta u_0^p$.}
   \State{Add fine-grid correction, $u_0 \leftarrow u_0 + \delta u_0^p$, in $\Omega_0^p$.}
   \columnbreak
   \\
   (mat-vec communication with $A_0$) \\
   (mat-vec communications with $R_l$) \\
   (communication with distance $\eta_l$ neighbors) \\
   (no communication) \\
   (no communication)
\end{multicols}
\end{algorithmic}
\end{algorithm}

Whereas AMG V-cycles require communication for each operation on each level of the hierarchy, AMG-DD requires communication only to calculate, restrict, and communicate the residual values between local AlgFAC solves. Algorithms \ref{vcyclealg} (AMG) and \ref{amgdd:alg} (AMG-DD) allow a high level comparison of the different communication requirements for each algorithm. The reduction in communication provided by AMG-DD comes with the added cost of redundant composite-grid storage and computation: any additional points in a processor's composite grid that were not in that processor's subdomain represent additional storage and computation cost. Due to the graded nature of the composite grids, this additional overhead may be kept to a manageable level (as studied in detail in Section \ref{results_sec}).


\subsection{Hybrid AMG-DD}

Up to this point, AMG-DD has been considered as an alternative cycling method to AMG V-cycles. It may be beneficial, however, to employ a hybrid method that uses regular AMG cycling on the finer levels and AMG-DD starting on a coarse level, $\tilde{L}$. As previously mentioned, it is common for the finest levels of an AMG hierarchy to be compute-limited while the coarser levels are communication-limited. This is due to the fact that the finer grids have simultaneously many more degrees of freedom and usually far smaller communication stencils than the coarse grids. Thus, the real need for communication reduction in an AMG cycle exists primarily (if not entirely) on the coarse levels. In a hybrid AMG-DD cycle, AMG-DD acts as a kind of coarse-grid solver for the AMG cycle, where the transition level, $\tilde{L}$, between AMG and AMG-DD is somewhere in the middle of the hierarchy. Algorithm \ref{hybridalg} shows this hybrid AMG-DD idea applied to a V-cycle, but this idea can be applied to other cycle structures as well, meaning that hybrid AMG-DD may enable more efficient versions of W or F cycles, where coarse-grid communication costs are even larger. Using AMG-DD in this way retains communication reduction on the coarse levels while also reducing the storage and computation overheads of the composite grids. 

Convergence rates for hybrid AMG-DD tend to be more similar to the convergence of the associated AMG V-cycle rather than the full AMG-DD algorithm. 
For the test problems shown in the numerical results in Section \ref{results_sec}, AMG-DD tends to converge faster than AMG V-cycles, making the hybrid scheme a less attractive option in this case. Furthermore, the GPU accelerated system used here provides sufficient local computational power that composite-grid overheads for the full AMG-DD algorithm are acceptable. Nevertheless, hybrid AMG-DD may be a useful option for problems where AMG-DD convergence suffers or for computing environments where local computation is a more significant cost and the computational overhead incurred by AMG-DD needs to be minimized.

\begin{algorithm}[H]
\caption{Hybrid AMG-DD V$(\nu_1,\nu_2)$-cycle}
\label{hybridalg}
\begin{algorithmic}
   \For {Downsweep: $l = 0,...,(\tilde{L}-1)$}
      \State{Relax $\nu_1$ times on $A_lu_l = f_l$.}
      \State{Restrict residual: $f_{l+1} \leftarrow R_l(f_l - A_lu_l)$}.
      \State{Initialize: $u_{l+1} \leftarrow \mathbf{0}$}.
   \EndFor
   \State{AMG-DD cycle: $u_{\tilde{L}} \leftarrow $ AMG-DD($A_{\tilde{L}}, f_{\tilde{L}}$) }.
   \For {Upsweep: $l = (\tilde{L}-1),...,0$}
      \State{Interpolate correction: $u_l \leftarrow u_l + P_lu_{l+1}$}.
      \State{Relax $\nu_2$ times on $A_lu_l = f_l$.}
   \EndFor
\end{algorithmic}
\end{algorithm}


\section{Algebraic fast adaptive composite cycling}
\label{fac_sec}

In the proof-of-concept implementation of AMG-DD used in \cite{Anonymous:KA9t5Fed}, the local composite-grid problems on each processor were solved by explicitly constructing composite-grid matrix operators and solving the resulting matrix equations with AMG. The additional overhead of forming the local composite-grid matrices (and subsequent associated AMG hierarchies) may be avoided, however, by cycling directly on the pieces of the original AMG hierarchy that make up the composite grid via an algebraic variant of the fast adaptive composite (FAC) method. FAC cycling has previously been applied to geometric multigrid hierarchies, and a detailed description of this can be found in \cite{McCormick:1989vx}. When attempting to apply the ideas behind FAC to an algebraic multigrid hierarchy, extra care must be taken in order to ensure proper calculation of values at the borders of each grid level. In the geometric case, these borders correspond to changes in the level of mesh refinement, and calculations there are based on information from the underlying discretization and mesh. In the algebraic setting, such geometric information is not available, and the AMG coarse grids may be quite irregular with arbitrary connectivity. Thus, this section discusses the necessary generalizations of previous geometric techniques for handling the borders of composite grids in order to achieve algebraic fast adaptive composite cycling (AlgFAC) on general AMG hierarchies. 

Classically, FAC cycles are used for geometric problems where high resolution of the grid is required only over some subset of the domain. A first step towards a more efficient algorithm for dealing with such a problem is to simply suppress relaxation where it is not necessary (i.e., outside of the region requiring high resolution). This saves the computational effort of relaxing where coarser representation of the solution is sufficient. The storage and computational cost can then be further reduced by removing the fine-grid representation of the solution where it is not needed altogether, resulting in a grid with varying levels of refinement similar to the AMG-DD composite grids. The resulting representation of the solution and multigrid cycling on this composite grid should be exactly equivalent to the algorithm that retains a fine-grid representation everywhere and simply suppresses relaxation. This will be the guiding principle used to develop AlgFAC cycling on algebraic grids: \emph{the result obtained via the AlgFAC cycle should be exactly the same as that obtained by cycling on the entire grid while suppressing relaxation at certain points}. This equivalence is shown in the following subsections by breaking the multigrid cycle down into its component parts: relaxation, interpolation, and restriction.

\subsection{Relaxation}

Denote the set of degrees of freedom that comprise grid level $l$ of the global AMG hierarchy as $\Omega_l$, and denote the composite-grid points on level $l$ as $D_l \subseteq \Omega_l$. The points in $D_l$ undergo relaxation as part of the AlgFAC cycle and are referred to as "real" points, whereas the remaining points on the level, referred to as "ghost" points, are not relaxed. In designing the AlgFAC cycling algorithm, the goal is to reproduce the action of a global AMG cycle (operating on the full $\Omega_l$ grids) where only points in $D_l$ are relaxed and as few ghost points as possible are actually stored or involved in any computation.

\begin{figure}
\centering
\includegraphics[width=0.7\textwidth]{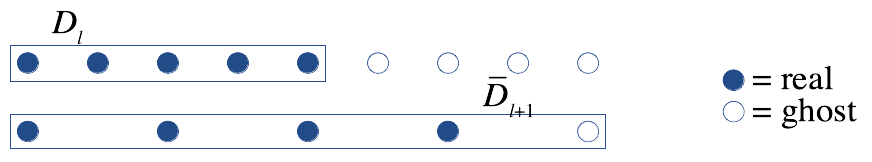}
\caption{Simple 1D illustration of real and ghost points and the sets $D_l$ and $\bar{D}_l$ on different levels (boxed points belong to the marked sets).}
\label{D}
\end{figure}

Applying simple, pointwise algebraic relaxation schemes like Jacobi or Gauss-Seidel to a real point, $i\in D_l$, requires solution values at all points, $j$, connected through the row stencil of $A_l$, i.e., a nonzero matrix entry, $a_{i,j}\in A_l$, connects points $i$ and $j$. Denote the extension of $D_l$ through the row stencil of $A_l$ as
\begin{align}
\bar{D}_l = D_l \cup \{j : a_{i,j}\in A_l,\,\, a_{i,j} \neq 0,\,\, i\in D_l\}.
\end{align}
Thus, in order to perform correct relaxation on each grid level, it is necessary to store solution values at all points in $\bar{D}_l$. Figure \ref{D} illustrates real and ghost points and the sets $D_l$, and $\bar{D}_{l}$ on different levels for an example composite grid using regular stencils on 1D grids.

\subsection{Interpolation}
\label{interpsec}

\begin{figure}
\centering
\includegraphics[width=0.7\textwidth]{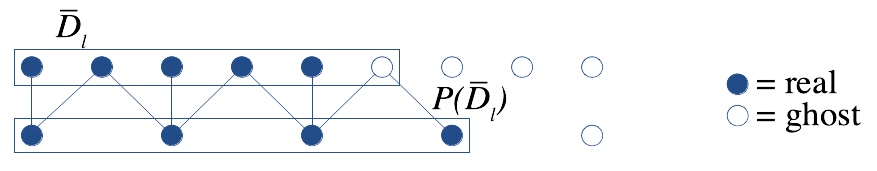}
\caption{Simple 1D illustration of the sets $\bar{D}_l$ and $P_l(\bar{D}_l)$ showing the connections to the coarse grid from $\bar{D}_l$ through the row stencil of $P_l$ (boxed points belong to the marked sets).}
\label{PD}
\end{figure}

\begin{figure}
\centering
\includegraphics[width=0.7\textwidth]{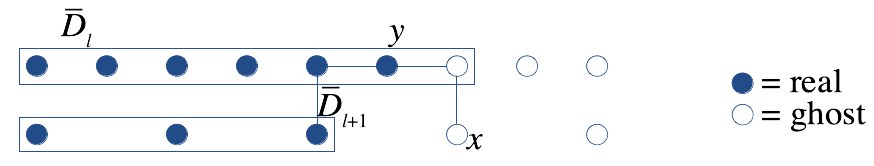}
\caption{Simple 1D illustration of the case where $x\in P_l(\bar{D}_l)$, but $x\not\in \bar{D}_{l+1}$ (boxed points belong to the marked sets).}
\label{PDninDlp1}
\end{figure}

Since relaxation requires correct solution values to be maintained at points in $\bar{D}_l$, it is necessary to interpolate coarse-grid corrections to these points. Denote the coarse-grid interpolatory set for $\bar{D}_l$ as $P_l(\bar{D}_l)$, i.e., points connected through the row stencil of $P_l$ as illustrated in Figure \ref{PD}. 
For typical AMG hierarchies, most points in $P_l(\bar{D}_l)$ are in $\bar{D}_{l+1}$ if $\eta_{l+1}\geq1$, so their contributions to interpolation will be correctly accounted for (note that in Figure \ref{PD} $P_l(\bar{D}_l)\subseteq\bar{D}_{l+1}$). 
It is possible, however, that some points in $P_l(\bar{D}_l)$ are outside of $\bar{D}_{l+1}$.
Consider, for example, the stencils drawn in Figure \ref{PDninDlp1}.
If the fine-grid point, $y$, has no connection through interpolation or restriction to the coarse grid (i.e. all entries in $P_l$ and $R_l$ associated with $y$ are zero) while its C-point neighbors are injected to and from the coarse grid as usual, then the coarse-grid point, $x$, is in $P_l(\bar{D}_l)$, but has no connection through $A_{l+1} = R_l^TA_lP_l$ to any real points on the coarse grid and is thus not in $\bar{D}_{l+1}$.
A point that is in $P_l(\bar{D}_l)$ but not in $\bar{D}_{l+1}$ is, by assumption, not a real degree of freedom and is thus initialized to zero and not relaxed on level $l+1$. The solution value at such a point will remain zero (and thus does not need to be accounted for in interpolation) so long as it also receives zero coarse-grid correction. That is, the set formed by recursively moving through the interpolatory sets of this point all the way to the coarsest level should not include any real degrees of freedom on any level. 
This is true for many hierarchies in practice including all test cases examined in this paper, but is not necessarily guaranteed in general. Assuming this property, all points in $P_l(\bar{D}_l)$ are either accounted for on the composite grid on the next coarse level or have zero contribution (and may thus be simply omitted), and interpolation may proceed as normal by applying $P_l$ to correction values on composite-grid level $l+1$.

\subsection{Traditional restriction}

As described above, the implementation for relaxation and interpolation over the composite grid is identical to that for a global multigrid cycle. Relaxation is simply not done for points outside $D_l$, and interpolation is not done for points outside $\bar{D}_l$. A similarly straightforward implementation of restriction presents problems, however. Since the coarse composite grids expand to cover the entire computational domain, storing the restriction stencils for all points in the composite grid in a recursive way such that it is always possible to restrict from fine to coarse grids in the standard way results in storing the entire global hierarchy. Thus, restriction must be somehow rewritten for composite-grid points whose restriction stencils do not reside in the composite grid.

\begin{figure}
\centering
\includegraphics[width=0.75\textwidth]{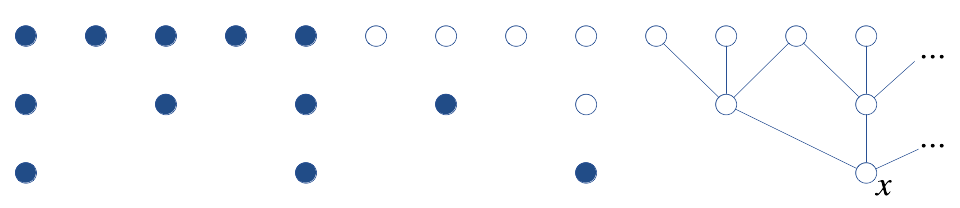}
\caption{Recursive restrictions stencils of a point, $x$, on a coarse level that remain separated by more than distance 1 from the real composite-grid points on finer levels.}
\label{restrictionStencils}
\end{figure}

A key realization is that actual inter-grid restriction is not necessary when finer-grid relaxation does not impact the residual. For the discussion here, assume the use of a V-cycle, though all ideas are easily extended to arbitrary cycle types. Consider restriction to composite-grid points whose restriction stencils include only points that are greater than distance 1 away from real points on all finer grid levels, e.g. the point, $x$, illustrated in Figure \ref{restrictionStencils}. Since corrections are initialized to zero on the downsweep and no relaxation occurs to change those values outside the real points, residual calculation simply yields the currently stored right-hand side for all coarse grids. Thus, setting the right-hand side, $f_{l+1,i}$, on level $l+1$ for V-cycle iteration $i$ by restricting the residual from level $l$ to level $l+1$ may be rewritten as a restriction of the finest-grid residual after the previous V-cycle iteration as follows:
\begin{align}
f_{l+1,i} &= R_lf_{l,i} \\
&= R_lR_{l-1}f_{l-1,i} \\
&\vdots \\
&= R_l...R_0(f_0 - A_0u_{0,i-1}).
\end{align}
Note that the fine-grid right-hand side, $f_{0,i} = f_0$, remains constant from iteration to iteration. Here, $u_{0,i-1}$ is the fine-grid solution at the end of the previous V-cycle iteration $i-1$. Again considering only composite-grid points on levels $0,...,l$ that are distance 1 away from real points means that $u_{0,i-1}$ is unaffected by relaxation on levels $0,...,l$ and may thus be expressed as the previous solution, $u_{0,i-2}$, plus the interpolated corrections from coarser levels generated on iteration $i-1$ all the way down to level $l$ as follows:
\begin{align}
f_{l+1,i} &= R_l...R_0(f_0 - A_0u_{0,i-1}) \\
&= R_l...R_0(f_0 - A_0(u_{0,i-2} + P_0u_{1,i-1})) \\
&= R_l...R_0(f_0 - A_0(u_{0,i-2} + P_0P_1u_{2,i-1})) \\
&\vdots \\
&= R_l...R_0(f_0 - A_0(u_{0,i-2} + P_0...P_lu_{l+1,i-1})).
\end{align}
Now, rearranging and invoking the definition of the coarse-grid operators as $A_{l+1} = R_lA_lP_l$ shows that restricting the residual may, in fact, be rewritten as a residual recalculation on the coarse grid using the previous solution/correction value and right-hand side as follows:
\begin{align}
f_{l+1,i} &= R_l...R_0(f_0 - A_0u_{0,i-2}) - R_l...R_0A_0P_0...P_lu_{l+1,i-1} \\
&= f_{l+1,i-1} - A_{l+1}u_{l+1,i-1}.
\end{align}
Thus, away from the real points where relaxation occurs, restricting the residual can be replaced with a residual recalculation on the coarse grid, removing the necessity for storing information on the grids above. Meanwhile, traditional restriction may occur as usual for points whose restriction stencils do include the influence of finer-grid relaxation. Note that this implementation still requires some additional layers of ghost points to be stored and computed on. Specifically, ghost points within distance 2 plus twice the interpolation distance are required, i.e., distance 4 or 6 neighbors when distance 1 or 2 interpolation is used, respectively. 

\subsection{AlgFAC restriction}

While the above method for implementing restriction is feasible, it still demands somewhat large overhead by requiring storage and computation at ghost points within distance 4 to 6 of the real points. Furthermore, initial right-hand side data must be communicated at all of these ghost points between each outer AMG-DD iteration, so the ghost points represent a significant overhead not only in terms of storage and computation but also in terms of communication cost. Restriction may actually be correctly performed without these additional ghost points, however, if the idea of coarse-grid residual recalculation is applied to \emph{all} points, even those influenced by finer-grid relaxation. This is how restriction is performed in an AlgFAC cycle. 

Rewriting restriction in this way requires some additional notation. Define $v_{l,i}$ as the value stored in the solution/correction vector just before restriction occurs from level $l$ to level $l+1$ on iteration $i$. Define $u_{l,i}$ as the value stored in the solution/correction vector just before interpolation occurs from level $l$ to level $l-1$ on iteration $i$. Respectively define $\delta^1_{l,i}$ and $\delta^2_{l,i}$ as the effects of pre- and post-relaxation on level $l$, iteration $i$. The following equations express the relationship between each of these terms during a V-cycle:
\begin{align}
v_{l,i} = \begin{cases}
u_{0,i-1} + \delta^1_{0,i}\,, & l=0 \\
\delta^1_{l,i}\,, & \text{else}
\end{cases} \\
u_{l,i} = v_{l,i} + P_lu_{l+1,i} + \delta^2_{l,i}.
\end{align}
The right-hand sides, $f_{l,i}$, on level $l$, iteration $i$ are then written as follows:
\begin{align}
f_{l,i} &= \begin{cases}
f_0\,, & l=0 \\
R_{l-1}(f_{l-1} - A_{l-1}v_{l-1,i})\,, &\text{else}.
\end{cases}
\end{align}
The goal is to reexpress restriction in a form that requires as little computation and storage as possible outside of the real degrees of freedom. Begin by rewriting restriction to the first coarse grid as follows:
\begin{align}
f_{1,i} &= R_0(f_{0,i} - A_0v_{0,i}) \\
&= R_0(f_0 - A_0(u_{0,i-1} + \delta^1_{0,i})) \\
&= R_0(f_0 - A_0(v_{0,i-1} + P_0u_{1,i-1} + \delta^2_{0,i-1} + \delta^1_{0,i})) \\
&= R_0(f_{0,i-1} - A_0v_{0,i-1}) - R_0A_0P_0u_{1,i-1} - R_0A_0(\delta^2_{0,i-1} + \delta^1_{0,i}) \\
&= f_{1,i-1} - A_1u_{1,i-1} - R_0A_0(\delta^2_{0,i-1} + \delta^1_{0,i}) .
\end{align}
Thus, restricting the residual to the first coarse grid may be written as a recalculation of a residual on the coarse grid using the currently stored right-hand side and correction vector, $f_{1,i-1} - A_1u_{1,i-1}$, plus an update term that restricts the effect of fine-grid relaxation, $- R_0A_0(\delta^2_{0,i-1} + \delta^1_{0,i})$.

It is possible to write restriction to the second coarse grid in a similar way as follows, recursively using the expression above for $f_{1,i}$ in order to rewrite $f_{2,i}$:
\begin{align}
f_{2,i} &= R_1(f_{1,i} - A_1v_{1,i}) \\
&= R_1(f_{1,i-1} - A_1u_{1,i-1} - R_0A_0(\delta^2_{0,i-1} + \delta^1_{0,i})  - A_1\delta^1_{1,i}) \\
&= R_1(f_{1,i-1} - A_1(v_{1,i-1} + P_1u_{2,i-1} + \delta^2_{1,i-1}) - R_0A_0(\delta^2_{0,i-1} + \delta^1_{0,i})  - A_1\delta^1_{1,i}) \\
&= R_1(f_{1,i-1} - A_1v_{1,i-1}) - R_1A_1P_1u_{2,i-1} - R_1A_1(\delta^2_{1,i-1} + \delta^1_{1,i}) - R_1R_0A_0(\delta^2_{0,i-1} + \delta^1_{0,i}) \\
&= f_{2,i-1} - A_2u_{2,i-1} - R_1A_1(\delta^2_{1,i-1} + \delta^1_{1,i}) - R_1R_0A_0(\delta^2_{0,i-1} + \delta^1_{0,i}) .
\end{align}
Thus, restriction to any level of the multigrid hierarchy may be rewritten as follows as a residual recalculation on the coarse level plus some restricted updates due to finer-level relaxation:
\begin{align}
f_{l,i} &= f_{l,i-1} - A_lu_{l,i-1} \\
& - R_{l-1}A_{l-1}(\delta^2_{l-1,i-1} + \delta^1_{l-1,i}) \\
& - R_{l-1}R_{l-2}A_{l-2}(\delta^2_{l-2,i-1} + \delta^1_{l-2,i}) \\
& \vdots \\
& - R_{l-1}R_{l-2} ... R_0A_0(\delta^2_{0,i-1} + \delta^1_{0,i}).
\end{align}
The restricted update terms above may be accumulated in a recursive way during the downsweep of the AlgFAC cycle such that obtaining the right-hand side on level $l$ involves only a single application of $R_l$ to a vector on level $l-1$ that has collected update terms from all finer levels.

\subsection{AlgFAC cycle}

With restriction rewritten as in the previous subsection, it is possible to write the AlgFAC V-cycle as a whole as shown in Algorithm \ref{facalg}. Auxiliary vectors, $t_l$ and $s_l$, are introduced in order to accumulate and restrict the effects of relaxation. For other cycle types, such as W or F cycles, a similar description is possible. The update vector, $t_l$, accumulates changes due to relaxation, $s_l$ accumulates updates from finer grids, and these updates are combined and passed to coarser grids during restriction, after which they are reset to zero. Rewriting things in this way requires only a single layer of ghost points and, furthermore, does not require right-hand side information there. Right-hand side information is needed at real points in order to perform relaxation and coarse-grid residual recalculation but is no longer required in the restriction stencils of real points. The update term, $t_l$, is nonzero only at real points, so $A_lt_l$ is nonzero only in $\bar{D}_l$. Some assumptions on the connectivity of the AMG hierarchy are required to ensure the stored composite grids correctly account for contributions to $s_l$ throughout the cycle. In fact, this requirement is very similar to the requirement for correct interpolation discussed in Section \ref{interpsec} and is equivalent to that requirement if $R_l=P_l^T$. Denote the restriction transpose stencil for $\bar{D}_l$ as $R^T_l(\bar{D}_l)$, i.e., points on level $l+1$ that are connected through the columns of $R$ to points in $\bar{D}_l$. Most points in $R^T_l(\bar{D}_l)$ will be naturally contained in $\bar{D}_{l+1}$, and contributions to $s_{l+1}$ will be accounted for at these points. If a point in $R^T_l(\bar{D}_l)$ is not in $\bar{D}_{l+1}$, then the value of $s_l$ is not needed here since no relaxation occurs and, subsequently, no right-hand side is needed. Moreover, the value of $s_l$ at this point need not be restricted to coarser levels, provided that moving recursively further through the restriction transpose stencils all the way to the coarsest level includes no real points. If $R_l = P_l^T$, then the restriction transpose stencils are exactly the interpolation stencils (or interpolatory sets), and this is the same requirement discussed in Section \ref{interpsec}.

\begin{algorithm}[H]
\caption{AlgFAC V-cycle}
\label{facalg}
\begin{algorithmic}
   \State{Initialize solution/corrections: $u_l \leftarrow \mathbf{0}, \forall l$}
   \State{Initialize updates: $t_l \leftarrow \mathbf{0}, s_l \leftarrow \mathbf{0}, \forall l$}
   \For {Iterate: $i = 0, 1,...$}
      \For {Downsweep: $l = 0,...,(L-1)$}
         \State{If $l\neq 0$, initialize: $u_l \leftarrow \mathbf{0}$}
         \State{Relax: $u_l \leftarrow u_l + \delta^1_l$}
         \State{Accumulate update: $t_l \leftarrow t_l + \delta^1_l$}
         \State{Restrict update: $s_{l+1} \leftarrow R_l(s_l + A_lt_l)$}
         \State{Recalculate coarse residual: $f_{l+1} \leftarrow f_{l+1} - A_{l+1}u_{l+1}$}
         \State{Subtract restricted update: $f_{l+1} \leftarrow f_{l+1} - s_{l+1}$}
         \State{Reset updates: $t_l \leftarrow \mathbf{0}, s_l \leftarrow \mathbf{0}$}
      \EndFor
      \State{Coarse solve: $u_L \leftarrow A_L^{-1}f_L$}
      \For {Upsweep: $l = (L-1),...,0$}
         \State{Interpolate: $u_l \leftarrow u_l + P_lu_{l+1}$}
         \State{Relax: $u_l \leftarrow u_l + \delta^2_l$.}
         \State{Accumulate update: $t_l \leftarrow t_l + \delta^2_l$}
      \EndFor
   \EndFor
\end{algorithmic}
\end{algorithm}

Compared to a standard AMG V-cycle, the AlgFAC V-cycle shown in Algorithm \ref{facalg} contains a few extra operations, but the extra computational cost of these operations is not too significant. Simple operations on each point that do not involve multiplication by a matrix (e.g. accumulation of the updates due to relaxation, or resetting vectors to zero) have small cost relative to mat-vecs. Most of the mat-vecs in Algorithm \ref{facalg} also occur in a standard AMG V-cycle (i.e., relaxation, interpolation, and restriction) with the same cost. The only additional mat-vec required in Algorithm \ref{facalg} is the coarse-grid residual recalculation, $f_{l+1} \leftarrow f_{l+1} - A_{l+1}u_{l+1}$. This additional mat-vec need not occur on the initial downsweep, however, since $u_{l+1}$ is initialized to zero, and it never happens on the finest grid, meaning that this is a relatively small extra computational expense.


\section{Residual communication algorithm}
\label{res_comm_sec}

Apart from the local AlgFAC solves, the other main piece of the AMG-DD algorithm that requires special attention is the residual communication algorithm by which processors receive right-hand side values at composite-grid points outside their subdomains. Denote processor $p$'s subdomain on level $l$ as $\Omega_l^p$ and its composite grid on level $l$ as $D_l^p$. Then, during the residual calculation and restriction steps of Algorithm \ref{amgdd:alg}, a given processor obtains updated residuals at points in $\Omega_l^p$, and then these updated values must be distributed to other processors that use these points in their composite grids, that is, processors $q$ such that $D_l^q\cap \Omega_l^p \neq \emptyset$. This is a non-trivial process, since the processor that owns the updated information does not have global information about which other processors' composite grids overlap its subdomain (and thus need to receive information from this processor). In addition, even if this information were available, a direct broadcast from the owning processor to all other processors that must receive information from that processor would result in inefficient communication with far too many messages being sent. Some kind of intelligent accumulation and distribution of information is needed, analogous to performing an efficient Allgather collective communication as opposed to naive point-to-point communication.

The AMG-DD residual communication algorithm was originally proposed (but not implemented) in \cite{Anonymous:KA9t5Fed}, and this paper provides empirical study of a parallel implementation of this algorithm in practice. The algorithm functions by recursively building up composite grids starting from the coarsest level of the AMG hierarchy and moving to the finest level. On each level, $l$, processors communicate with distance $\eta_l$ processor neighbors, where $\eta_l$ is the padding on level $l$ and distance is measured through the graph of $A_l$. It is assumed that distance 1 processor neighbors are known, since this information is required to perform mat-vecs with $A_l$ (and if this information is not available, algorithms exist to efficiently obtain it \cite{Anonymous:ycC62NK_}). From this starting point, arbitrary-distance neighbors may be found by recursively communicating with processor neighbors at longer and longer distances (growing the set of processor neighbors with each communication). 

Once distance $\eta_l$ processor neighbors are known, the AMG-DD residual communication algorithm proceeds as follows. Beginning on the coarsest level, $L$, each processor, $p$, with a non-empty subdomain (note that if the coarsest level has fewer points than there are processors, some processors will remain inactive here) communicates with each of its distance $\eta_L$ processor neighbors, $q$, sending residuals at points in the set, $\Psi = \{i\in \Omega_L^p : dist(i, \Omega_L^q) \leq \eta_L\}$, to processor $q$, where $dist(\cdot,\cdot)$ measures distance through the graph of $A_l$. This routine then repeats on the next finer level, $L-1$, but in addition to points in the set $\Psi= \{i\in \Omega_{L-1}^p : dist(i, \Omega_{L-1}^q) \leq \eta_{L-1}\}$, each processor, $p$, sends a composite-grid structure based on $\Psi$ to processor $q$. More precisely, take the $C$-points of $\Psi$ on the next coarse level, add points within the padding distance, and call the resulting set $\Psi_c$. Then residuals for all points in $\Psi\cup\Psi_c$ are communicated. After this communication, each processor has updated residuals in a composite grid based on its subdomain from level $L-1$ down. The recursion is now apparent. The same process is repeated moving up to the finest level, with each processor sending distance $\eta_l$ points, $\Psi$, on level $l$ plus a composite-grid structure, $\Psi_c$, containing points on coarser levels. By recursive argument, after communication on each level, $l$, all processors have correctly updated residual values in a composite grid based on their subdomain on level $l$. A more detailed proof of the correctness of this algorithm as well as some further illustration of the communicated $\Psi$ and $\Psi_c$ sets are given in \cite{Anonymous:KA9t5Fed}. Pseudocode for this residual communication algorithm is given in Algorithm \ref{res_comm:alg}

\begin{algorithm}[t]
\caption{AMG-DD residual communication algorithm}
\begin{algorithmic}
  \For {Loop over levels: $l=L,...,0$ }
      \For {Loop over distance $\eta_l$ processor neighbors: $q=1,... $ }
         \State { Let $\Psi = \{i\in \Omega_l^p : dist(i, \Omega_l^q) \leq \eta_l\}$ }
         \State { Form a composite grid $\Psi_c$ based on $\Psi$ }
         \State { Send residuals at points in $\Psi\cup \Psi_c$ to processor $q$ }
      \EndFor
  \EndFor
\end{algorithmic}
\label{res_comm:alg}
\end{algorithm}

Note that the communication stencils used for the residual communication algorithm are based on the padding, $\eta_l$, used on each level. In the case where $\eta_l = 1$ $\forall l$, the communication stencils are exactly those used for mat-vecs on each level with $A_l$ and the $\Psi$ sets exactly comprise the mat-vec halo points (message sizes will be larger due to the need to send the $\Psi_c$ sets as well). Using larger $\eta_l$ may result in expanded communication stencils and subsequently more overall messages as well as a larger total volume of data being communicated, as shown in more detail in Section \ref{results_sec}. On finer levels, reaching further through the graph of $A_l$ may not jump to more distant processors, but on coarse levels this is likely: on a coarse level where each processor has single degree of freedom, for example, any increase to $\eta_l$ will expand the set of processor neighbors, potentially leading to a significant increase in the total number of messages sent. This highlights the importance of rewriting restriction in the AlgFAC cycle as described in Section \ref{fac_sec}. The more straightforward approach to restriction demands right-hand side information at multiple layers of ghost points (typically 4 to 6 layers), which in turn demands that the residual communication algorithm communicate information at distance $\eta_l$ \emph{plus} the number of ghost layers, resulting in significantly more messages and total volume sent. By rewriting restriction, however, right-hand side information is only needed at real degrees of freedom, enabling the residual communication algorithm to communicate information only at the distance of the padding as described above.

Note also that the residual communication algorithm does not require processors to have global information about off-processor composite grids (only distance $\eta_l$ processor neighbors and corresponding distance $\eta_l$ points are needed), nor does it even require a given processor to know the size and shape of its own composite grid a priori. Thus, this algorithm may also be used to setup the composite grids in an efficient manner. The top down process of expanding by the padding, coarsening, and repeating is a convenient way to define composite grids, but is not an efficient way to construct them in practice in a distributed setting. Thus, the residual communication algorithm may be employed to construct the composite grids simply by passing matrix information (rows of $A_l$, $P_l$, and $R_l$) instead of residual values. Note that this does require communicating information at distance $\eta_l + 1$ on each level in order to obtain the matrix information at the ghost points. 

One important caveat about the residual communication algorithm is that it communicates a significant amount of redundant information if implemented as described above. This redundancy has two causes. First, processor $p$ may communicate information at some point on level $l$ to processor $q$ and then send the same information to the same destination again later if that point is included in a $\Psi_c$ composite-grid structure being sent to processor $q$ on some finer level. Second, it is possible that two processors $p$ and $r$ have accumulated updated information at some point and both processors then send information at this point to processor $q$. The first source of redundancy is easily rectified, since the redundant information is sent from the same processor: when constructing a $\Psi_c$ composite-grid structure to send, the sending processor may simply remove points previously sent to the same destination. The second source of redundancy is less trivial and requires a second round of communication during the setup phase, during which each processor sends back a list of redundantly received points to all processors it received from. This extra communication step need only happen once during the setup phase, however, and then all redundancy is avoided during the solve phase residual communications. 


\section{Numerical results}
\label{results_sec}

The model problems used throughout this section are variations on the diffusion problem,
\begin{align}
-\nabla \cdot K \nabla u &= f\,, \hspace{1 cm} \Omega \\
u &= 0\,, \hspace{1 cm} \partial \Omega.
\end{align}
The equation is discretized using linear $H^1$ conforming finite elements on regular meshes, using the MFEM\footnote{\url{mfem.org}} library to perform the discretization and METIS\footnote{\url{http://glaros.dtc.umn.edu/gkhome/views/metis}} to partition the problem among processors. 
Two versions of the problem are considered: 3D Poisson (denoted Problem 1 below), where $K = I$ and $\Omega$ is the unit cube; and 2D rotated anisotropic diffusion (denoted Problem 2 below), where $\Omega$ is the unit square, $K = Q^TDQ$, and $Q$ and $D$ are the rotation and scaling matrices respectively defined by
\begin{align}
Q &= \begin{bmatrix}
\cos(\theta) & \sin(\theta) \\
-\sin(\theta) & \cos(\theta)
\end{bmatrix},
&
D &= \begin{bmatrix}
1 & 0\\
0 & \epsilon
\end{bmatrix},
\end{align}
with $\theta = 3\pi/16$ and $\epsilon = 0.001$. A random initial guess and zero right-hand side are used throughout. These problems are chosen because they are well-known to have significant growth in communication stencils on the coarse levels of classical AMG hierarchies, leading to significant communication costs. 

All implementation of AMG-DD discussed in this paper is done in \emph{hypre}\footnote{\url{https://github.com/hypre-space/hypre} The AMG-DD code used for this paper can be found at commit hash c38c9cc of the hypre-space repository and will soon be included in the master branch.
} on top of BoomerAMG.
Parameter choices for the underlying AMG hierarchy as well as choices about the decomposition of the problem among processors all influence the size of the composite grids and subsequently the storage and computation overhead and overall performance of AMG-DD. Since the composite grids are constructed based on movement through the graph of $A_l$ on each level, the complexity of these coarse-grid operators has a direct impact on the size of the composite grids. Classical AMG and its initial extensions to the parallel setting generally yield higher complexity than current "best practices" for AMG \cite{DeSterck:2008fc, DeSterck:2006et}. Significantly larger composite grids result from Falgout coarsening and classical modified interpolation (classical AMG) compared to HMIS coarsening and extended + i interpolation (best practices AMG), which are designed to reduce coarse-grid complexity. The tests shown here use the latter settings for best practices (along with \emph{hypre}'s default parameters for strength of connection, interpolation truncation, etc.).

The size of the composite grids relative to the size of the subdomain on a processor also depends on the partitioning of the problem. Assuming the partitioning of the degrees of freedom among processors corresponds to a spatial partitioning of the underlying problem geometry, the size of a processor's composite grid is roughly dependent on the surface area of its subdomain. Thus, the overhead of the composite grids depends on the surface-area-to-volume ratio of the processor subdomains. Algebraically, this is simply the percentage of the degrees of freedom that have connections through the graph of the matrix to points on other processors. The dimensionality of the underlying problem as well as the number of degrees of freedom per processor both impact the surface-area-to-volume ratio: higher dimensional problems and a smaller number of degrees of freedom per processor both result in larger surface-area-to-volume ratio and, consequently, larger overheads for AMG-DD, as shown throughout the results below.

The use of GPU acceleration is increasingly common in modern supercomputers, and provides a promising computational environment for AMG-DD due to the relatively high cost of communication compared to computation. The fine-grained parallelism provided by the GPU enables very fast local computation, but also limits the choice of available smoothers. A typical smoother for parallel AMG is some form of hybrid Gauss-Seidel (this is the default in \emph{hypre}), but the sequential nature of the on-processor computation in these smoothers mean that they are not viable on the GPU. Simple Jacobi smoothers, while easily and efficiently implemented on the GPU, do not provide adequate smoothing for many problems. Thus, all results below use a more robust variant of Jacobi, a CF $l_1$-scaled Jacobi smoother \cite{Baker:2011ib}, that both runs efficiently on the GPU and retains good smoothing properties. The cycle type used is a V(1,1)-cycle, i.e. one pre- and one post-smoothing on each level. Convergence results and other statistics gathered on AMG-DD shown below were generated on the bwForCluster MLS\&WISO\footnote{\url{https://wiki.bwhpc.de/e/Category:BwForCluster_MLS&WISO_Development}} and timing results were obtained on the large GPU cluster, Piz Daint
\footnote{\url{https://www.cscs.ch/computers/piz-daint/}}.


\subsection{AMG-DD parameter choice}
\label{amgdd_params_sec}

AMG-DD introduces a few new important parameters to choose (on top of the many parameter choices involved in setting up the underlying AMG hierarchy for a given problem). One main question of interest, however, is how accurately each processor's composite-grid solution represents the desired correction to the global solution. There are two sources of error here: that due to the approximation of the global fine grid with a composite grid; and that due to inexact solution of the composite problem. The two primary parameters that affect these sources of error are the padding chosen in constructing the composite grids and the number and type of AlgFAC cycles used to solve on the composite grid between each global residual recalculation. The use of larger padding grows the overall size of the composite grid on each processor, allowing for the composite problem to more accurately represent the global problem but also incurring the additional cost of communicating and computing on more composite-grid degrees of freedom. Performing additional AlgFAC cycles between each global residual recalculation should generate solutions on each processor that are more accurate to their corresponding composite problems at additional computational cost but at no additional communication cost.

\begin{figure}
\centering
\includegraphics[height=0.3\textwidth]{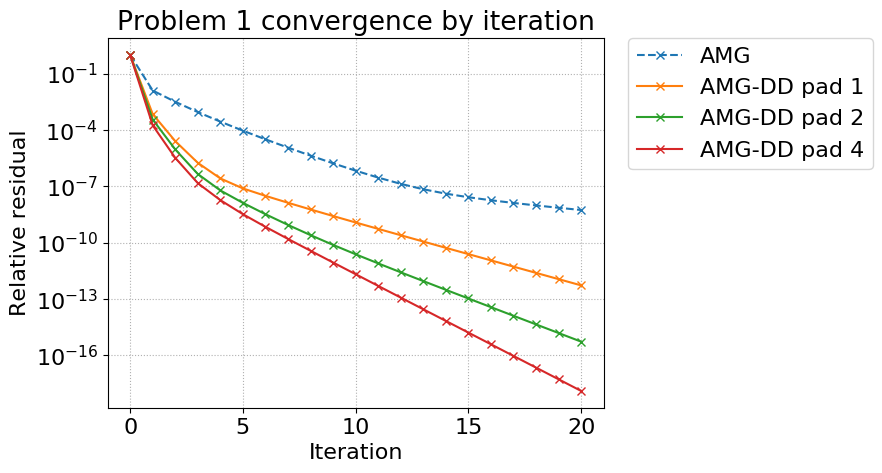}
\includegraphics[height=0.3\textwidth]{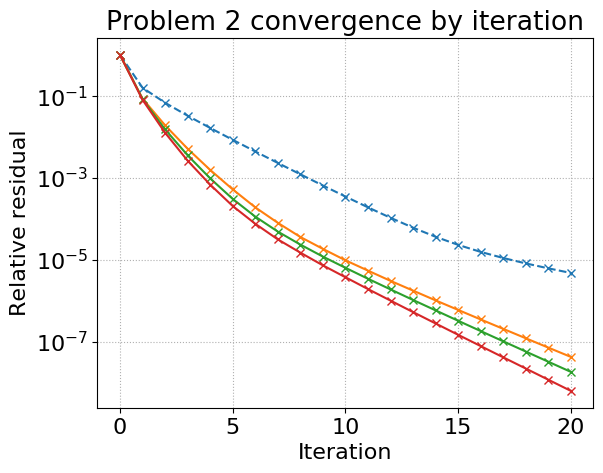}
\caption{Relative residual convergence for AMG vs. AMG-DD with various paddings for 3D Poisson (left) and rotated anisotropic diffusion (right).}
\label{resConvPad}
\end{figure}

Figure \ref{resConvPad} shows convergence of the relative residual by iteration for AMG vs. AMG-DD with different levels of padding for each test problem with 256 processors and approximately 100,000 degrees of freedom per processor on the fine grid for the 3D Poisson problem and the rotated anisotropic problem, respectively. Here, the same padding is chosen on each level, and the local composite problems on each processor are solved accurately (using 10 AlgFAC V-cycles) in order to isolate the effect of padding on accuracy of the composite problems to the global problem. As shown in Figure \ref{resConvPad}, larger paddings yield better AMG-DD convergence, as expected, but recall that larger paddings also incur larger costs both in terms of communication and computation. 

\begin{figure}
\centering
\includegraphics[height=0.3\textwidth]{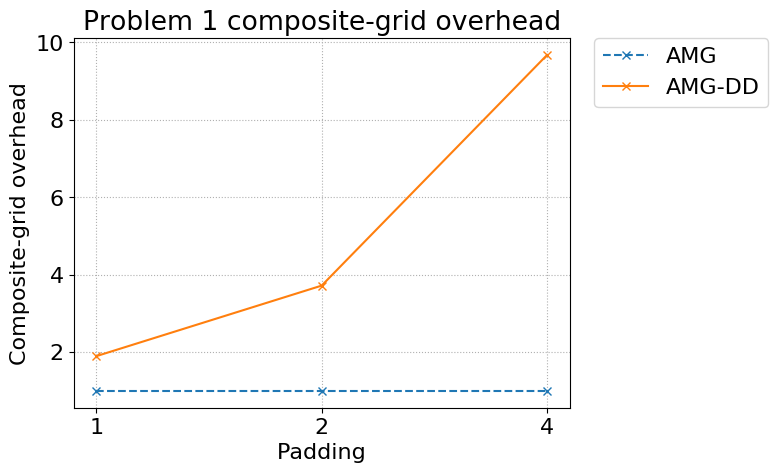}
\includegraphics[height=0.3\textwidth]{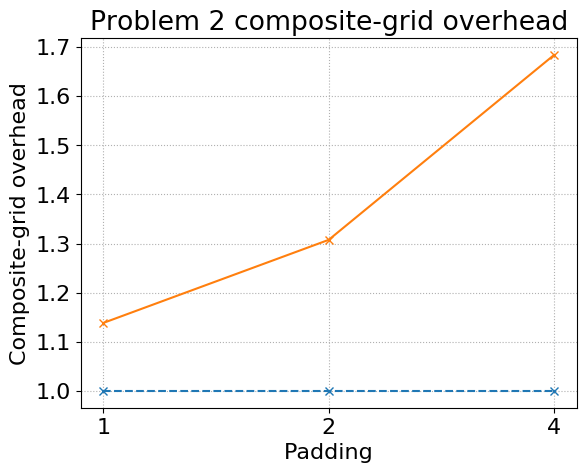}
\caption{Composite-grid overheads for AMG-DD as padding increases for 3D Poisson (left) and rotated anisotropic diffusion (right).}
\label{gridSizePad}
\end{figure}

Figure \ref{gridSizePad} shows the overhead incurred by the AMG-DD composite grids with different paddings compared to AMG.
Throughout this section, "composite-grid overhead" is defined as the total number of nonzeros of $A_l$ for the real degrees of freedom on all levels, $l$, stored across all composite grids (i.e., accounting for points stored redundantly in many composite grids) normalized by the total number of nonzeros stored in AMG (i.e., with no redundant storage). The number of nonzeros in $A_l$ serves here as a proxy for storage and computational costs of each method. As shown in Figure \ref{gridSizePad}, increasing the padding dramatically increases the storage and computational cost of AMG-DD relative to AMG, with significantly worse overheads in 3D due to higher surface-area-to-volume ratios for each processor's subdomain. 

\begin{figure}
\centering
\includegraphics[height=0.3\textwidth]{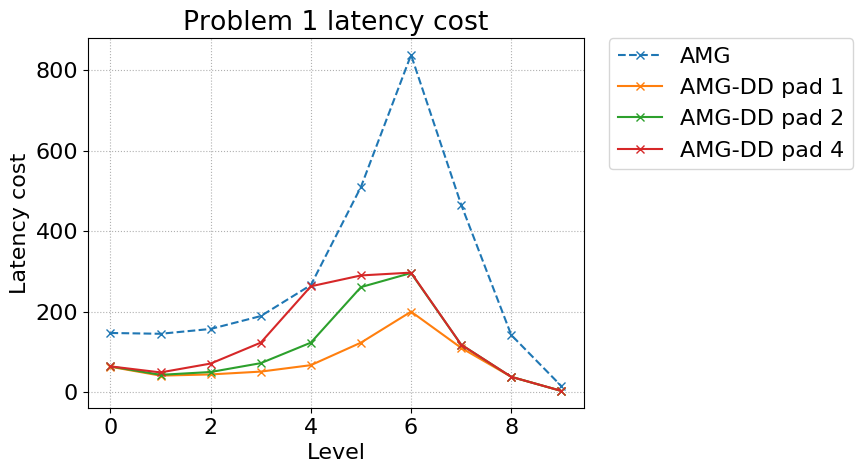}
\includegraphics[height=0.3\textwidth]{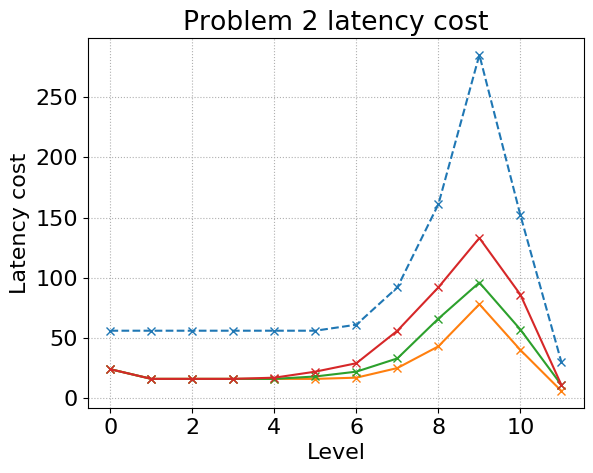}

\includegraphics[height=0.3\textwidth]{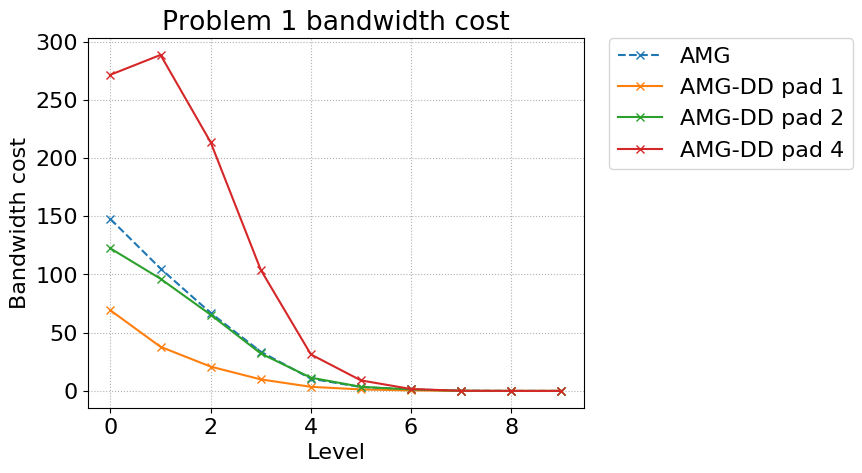}
\includegraphics[height=0.3\textwidth]{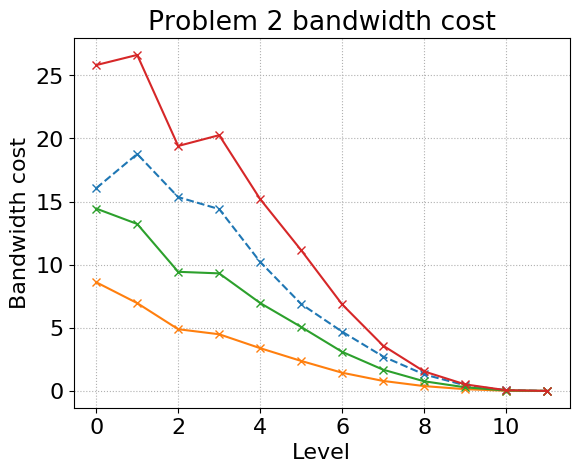}
\caption{Latency costs (top) and bandwidth costs (bottom) on each level for AMG vs. AMG-DD for different paddings for 3D Poisson (left) and rotated anisotropic diffusion (right).}
\label{commCostPad}
\end{figure}

Communication costs can also grow significantly with increased padding, as shown in Figure \ref{commCostPad}. 
Throughout this section, "latency costs" are measured as the sum over each communication stage for one iteration of the maximum number of messages sent from a single processor, and "bandwidth costs" are measured as the total volume of data (in MB) communicated over one iteration. 
Recall that for AMG, communication costs are the result of mat-vecs on each level: two mat-vecs with $A_l$ for each relaxation plus one for a residual calculation (for a total of five mat-vecs with $A_l$) and single mat-vecs with $P_l$ and $R_l$ for interpolation and restriction. AMG-DD must also do a mat-vec with $A_0$ on the finest grid as well as mat-vecs with $R_l$ on each level to restrict the residual down through the hierarchy, but the dominant communication cost for AMG-DD is the residual communication algorithm, Algorithm \ref{res_comm:alg}.
The highest volume of data is communicated on the fine levels, and, for larger paddings, AMG-DD communicates significantly more data due to the need to communicate deeper halos. The increase in the number of messages on coarse grids for AMG is observed here and mirrored by AMG-DD. Recall that the AMG-DD residual communication algorithm (which is the dominant communication cost) uses the same communication stencils as a mat-vec with $A_l$ on each level in the padding 1 case, but these communication stencils expand with larger paddings.
The growth in the number of messages for AMG-DD with higher padding may be alleviated in some cases by replacing coarse-level point-to-point communication with a collective Allgather of the global coarse grid (assuming a grid coarse enough that it is reasonable to redundantly store on all processors) or by grouping processors together on coarse levels, treating the group as a single rank in the residual communication algorithm, and doing some additional local collective communications. While these techniques can limit the growth in the number of messages for AMG-DD with higher padding, higher padding always incurs additional messages, and the volume of data communicated cannot be reduced.

\begin{figure}
\centering
\includegraphics[height=0.3\textwidth]{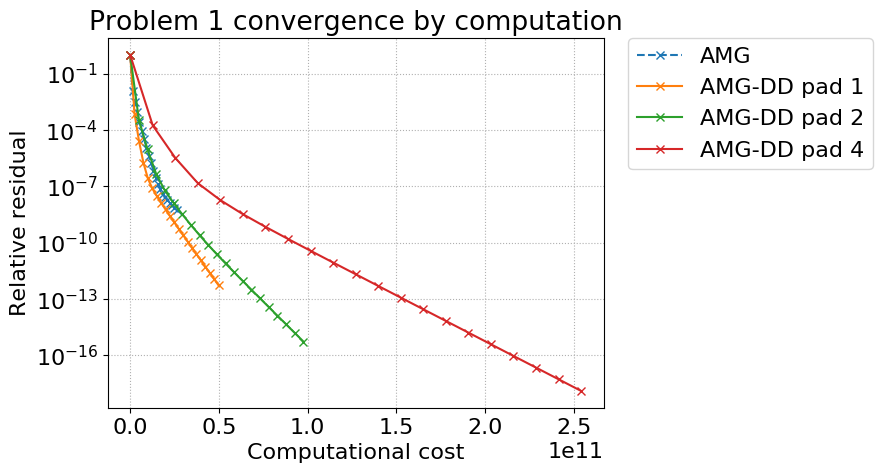}
\includegraphics[height=0.3\textwidth]{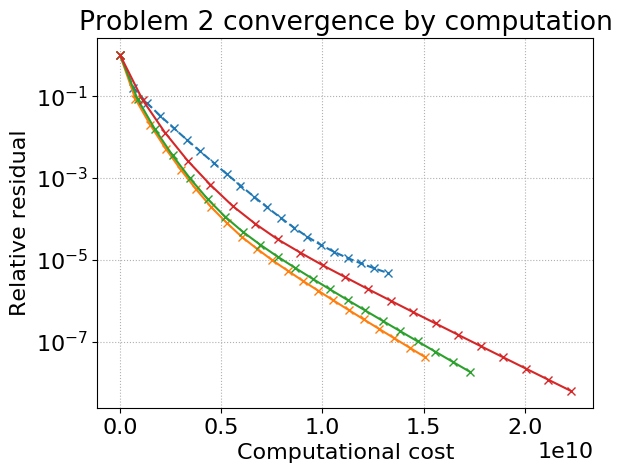}

\includegraphics[height=0.3\textwidth]{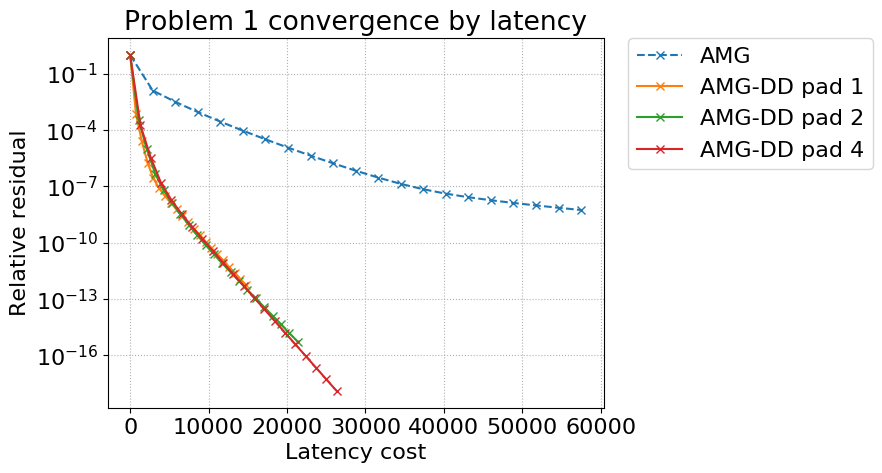}
\includegraphics[height=0.3\textwidth]{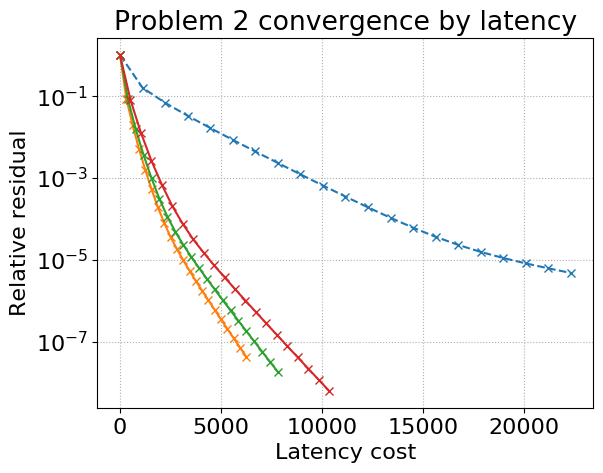}

\includegraphics[height=0.3\textwidth]{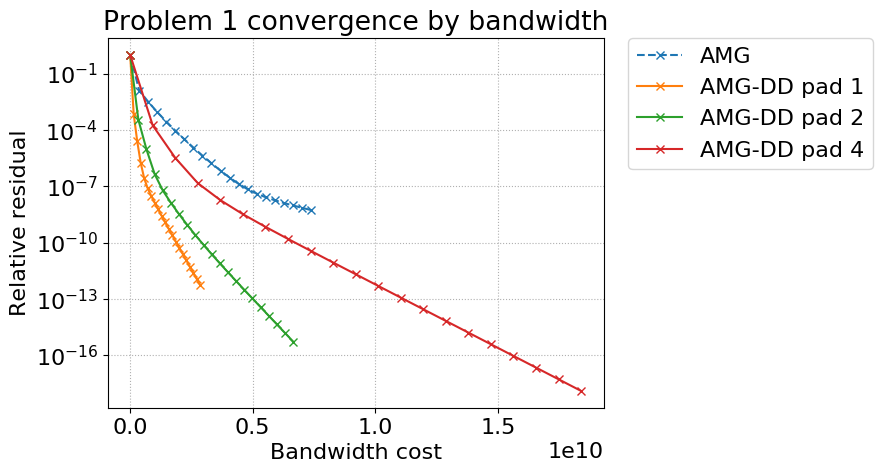}
\includegraphics[height=0.3\textwidth]{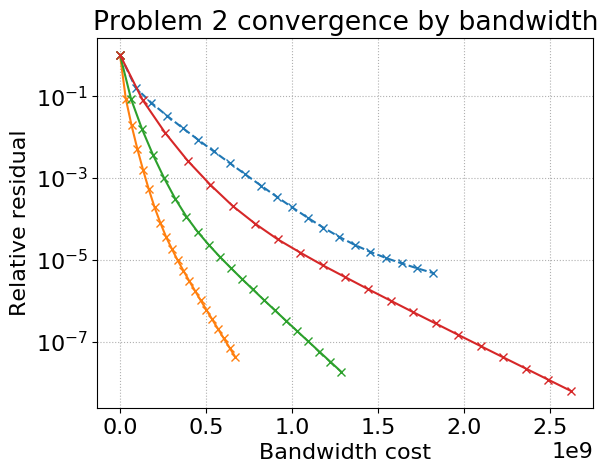}
\caption{Residual convergence by computational cost (top), latency cost (middle), and bandwidth cost (bottom) for AMG vs. AMG-DD for different paddings for 3D Poisson (left) and rotated anisotropic diffusion (right).}
\label{resConvByCost}
\end{figure}

For the test problems shown here, the convergence benefit generally does not justify the cost incurred by increased padding for AMG-DD. Figure \ref{resConvByCost} combines the convergence and cost results described above by plotting the residual reduction per cost, where, again, the costs are measured by the total number of nonzeros in $A_l$ for all $l$ and all processors (computational cost), the maximum number of messages sent by one processor summed over each communication stage (latency cost), and the total volume of data communicated (bandwidth cost). As shown, AMG-DD with padding 1 achieves better residual convergence with less cost compared to higher paddings. Also, AMG-DD with padding 1 achieves significantly better accuracy per communication cost than AMG.

\begin{figure}
\centering
\includegraphics[height=0.3\textwidth]{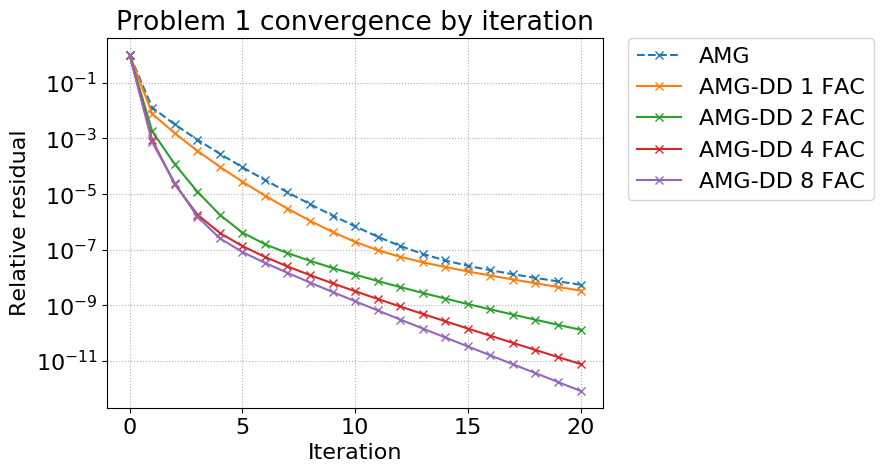}
\includegraphics[height=0.3\textwidth]{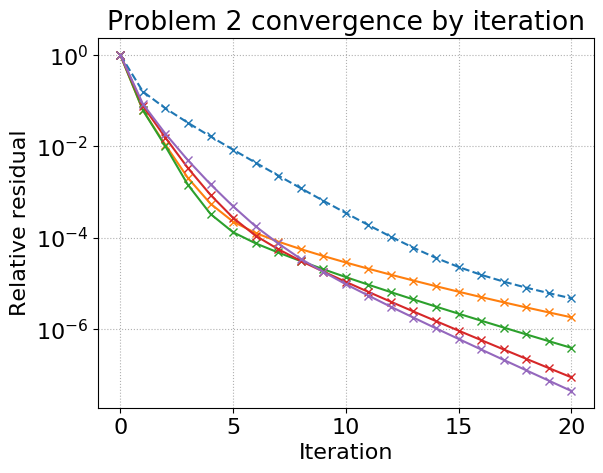}

\includegraphics[height=0.3\textwidth]{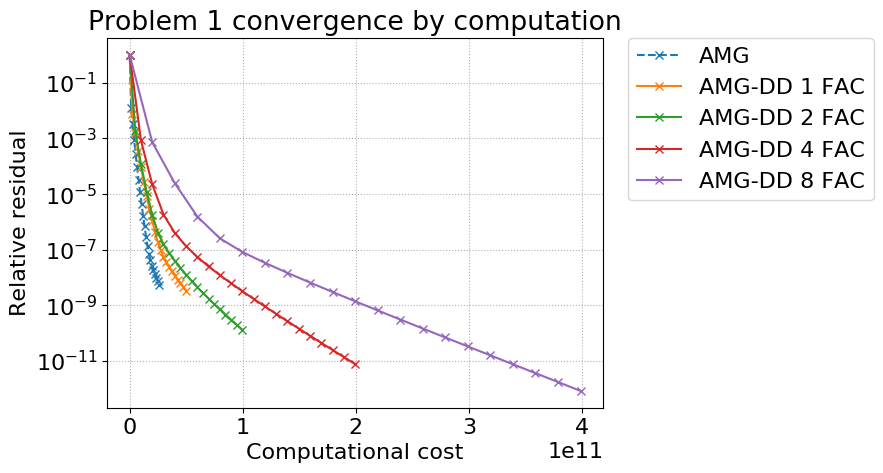}
\includegraphics[height=0.3\textwidth]{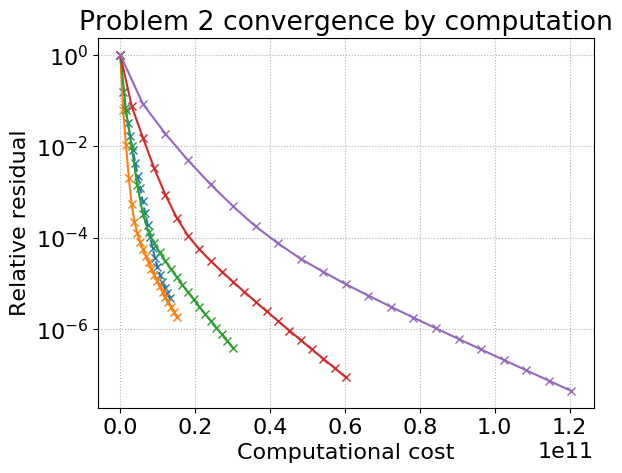}
\caption{Residual convergence by iteration (top) and computational cost (bottom) for AMG vs. AMG-DD for different paddings for 3D Poisson (left) and rotated anisotropic diffusion (right).}
\label{numFacFig}
\end{figure}

The number of inner AlgFAC cycles per AMG-DD iteration determines how well the composite grid problems are solved on each problem. While additional AlgFAC cycles incur no additional communication cost, they do come at additional computational expense. Thus, it is desirable to optimize the number of AlgFAC cycles in terms of accuracy per computational cost, where accuracy is measured by outer AMG-DD convergence rather than AlgFAC convergence to the composite solution. Figure \ref{numFacFig} shows residual convergence by iteration and by computational cost for AMG vs. AMG-DD with padding 1 and different numbers of AlgFAC V(1,1)-cycles. For the problems here, AMG-DD convergence begins to show diminishing returns beyond 4 AlgFAC cycles, indicating that the inner problem has sufficiently converged. More accurate composite solutions (obtained through more AlgFAC cycles) generally yield better AMG-DD convergence, but interestingly, for the rotated anisotropic diffusion problem, initial convergence can actually be better with less accurate composite solves, while asymptotic convergence is better with more accurate composite solves. For both problems, however, the additional computational cost of multiple AlgFAC cycles outweighs any benefits to convergence. AMG-DD achieves the best efficiency in terms of accuracy per computational cost for a single AlgFAC cycle and achieves computationally efficiency comparable to AMG in that case. If the computer architecture has very fast local computation compared with communication (as on GPU accelerated clusters), however, it may still be beneficial to do multiple AlgFAC cycles per AMG-DD iteration if the goal is to minimize time to solution.

\subsection{Weak scaling results}
\label{weak_scaling_sec}

With the above discussion on parameter choices in mind, weak scaling results comparing AMG and AMG-DD with padding 1 and two AlgFAC V(1,1) cycles for the composite solve are presented below. Timing results were obtained on the large GPU cluster, Piz Daint, using up to 1,024 GPUs. The local computational acceleration provided by the GPUs makes the use of multiple AlgFAC cycles per AMG-DD iteration a practical choice. For these weak scaling studies, the number of degrees of freedom is increased to 250,000 per processor.

\begin{figure}
\centering
\includegraphics[height=0.3\textwidth]{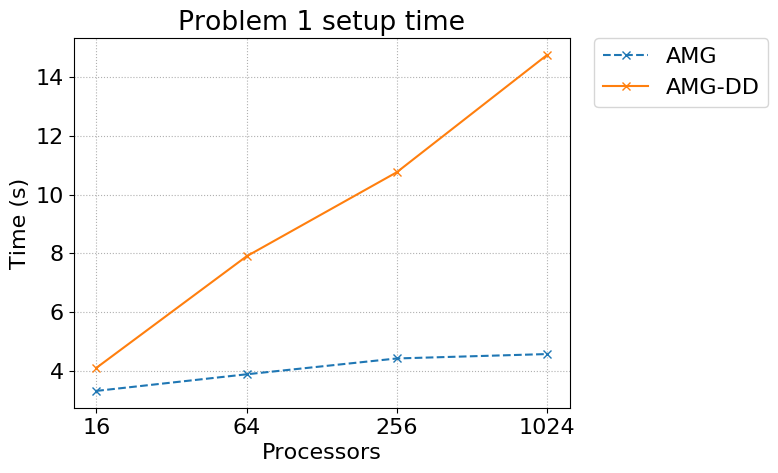}
\includegraphics[height=0.3\textwidth]{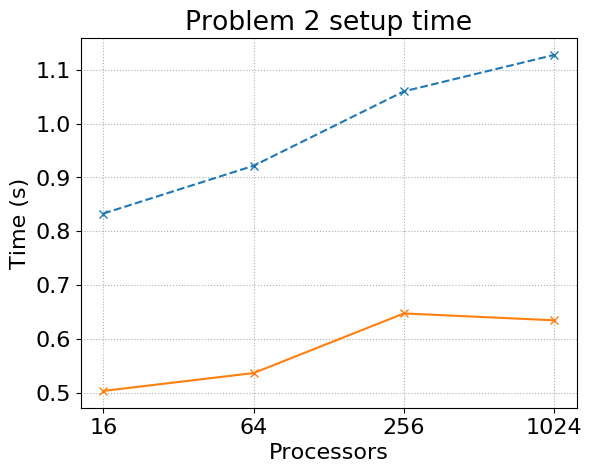}

\includegraphics[height=0.3\textwidth]{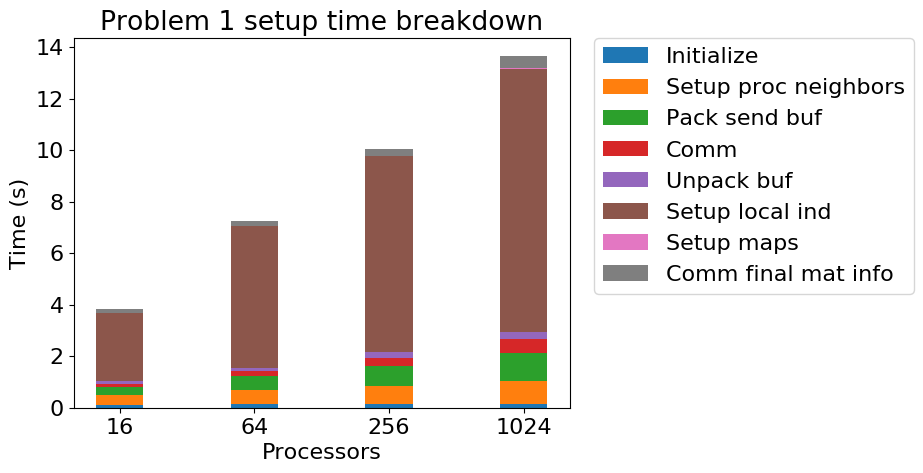}
\includegraphics[height=0.3\textwidth]{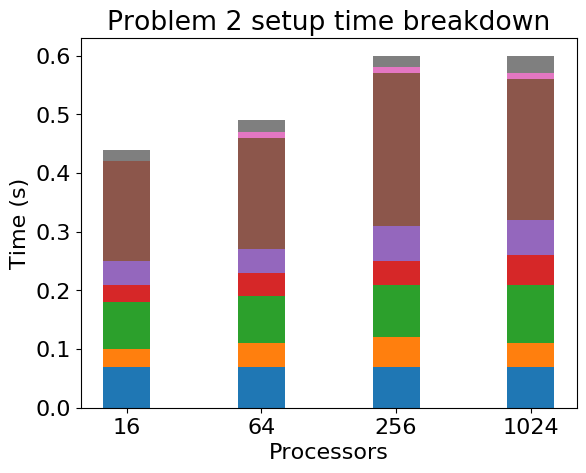}

\caption{AMG and AMG-DD setup time weak scaling and AMG-DD setup time breakdown for 3D Poisson (left) and rotated anisotropic diffusion (right).}
\label{setup}
\end{figure}

First, it is necessary to consider the setup cost of AMG-DD. In order for AMG-DD to be a viable alternative to AMG in practice, it must not introduce too much additional setup cost. From an algorithmic standpoint, the AMG-DD setup phase should be significantly cheaper than the AMG setup phase. For the AMG setup, the dominant cost is typically the triple matrix product used to generate the coarse-grid operators, $A_l$ on each level, $l$. The extra work of the AMG-DD setup then simply involves moving and reorganizing the resulting matrix data using communication stencils determined by the padding (recall that the setup requires communication of information for ghost points, so the communication stencils are based on distance $\eta_l+1$). In practice, however, searching through the composite grids to generate the $\Psi_c$ sets discussed in Section \ref{res_comm_sec} and the reindexing involved in incorporating received data to build up each processor's composite grid are non-trivial tasks to implement in an efficient manner. The code used for this paper is still under development, and optimization of the setup phase is ongoing. Nevertheless, current timings and profiling of the setup phase show promising results. Figure \ref{setup} shows weak scaling results for the setup time of AMG-DD vs. AMG as well as a rough breakdown of major costs for the AMG-DD setup. In the 2D case, the AMG-DD setup time scales well and is less than that required for AMG. In 3D, the AMG-DD setup time is much worse, but the setup time breakdown shows a clear dominant cost: setting up local indices for information received from off-processor (primarily mapping global to local column indices for the matrices $A_l$ on each level). This provides a clear path forward for the currently ongoing optimization of the AMG-DD code.
Note also that, at the time of this paper, the AMG setup implemented in \emph{hypre} has not been ported to the GPU, and currently \emph{none} of the AMG-DD setup routines utilize the GPU. Much of the local computational work in the AMG-DD setup phase can be efficiently ported to the GPU (including the dominant local reindexing portion of the code), which should provide a significant boost to setup performance.

\begin{figure}
\centering
\includegraphics[height=0.3\textwidth]{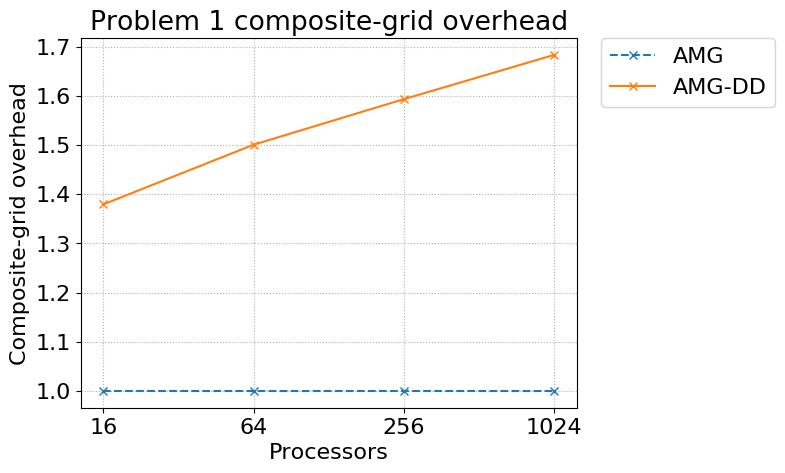}
\includegraphics[height=0.3\textwidth]{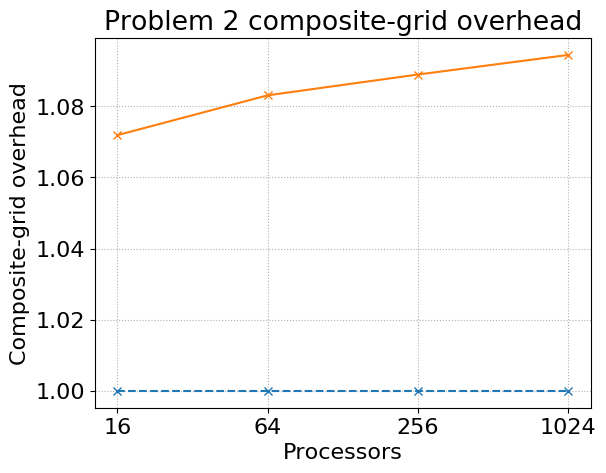}

\caption{Composite-grid overhead weak scaling for AMG-DD for 3D Poisson (left) and rotated anisotropic diffusion (right).}
\label{gridSizeWeakScaling}
\end{figure}

Next, some of the quantities of interests from Section \ref{amgdd_params_sec} are examined in a weak scaling context. Figure \ref{gridSizeWeakScaling} shows how the composite-grid overhead changes with the number of processors. 
Again, the overhead here is defined as the number of nonzeros of $A_l$ for the composite grid real degrees of freedom on all levels, $l$, divided by the total number of nonzeros of the $A_l$'s stored by AMG.
As noted above, these results use more degrees of freedom per processor compared to the results in Section \ref{amgdd_params_sec}, so AMG-DD achieves slightly smaller relative overhead in terms of storage and computation, and the growth of this overhead with the number of processors is slow (note the logarithmic scale for the processor axis), especially in the 2D case. This suggests that overheads should remain manageable when scaling AMG-DD to even larger processor counts. 

\begin{figure}
\centering
\includegraphics[height=0.3\textwidth]{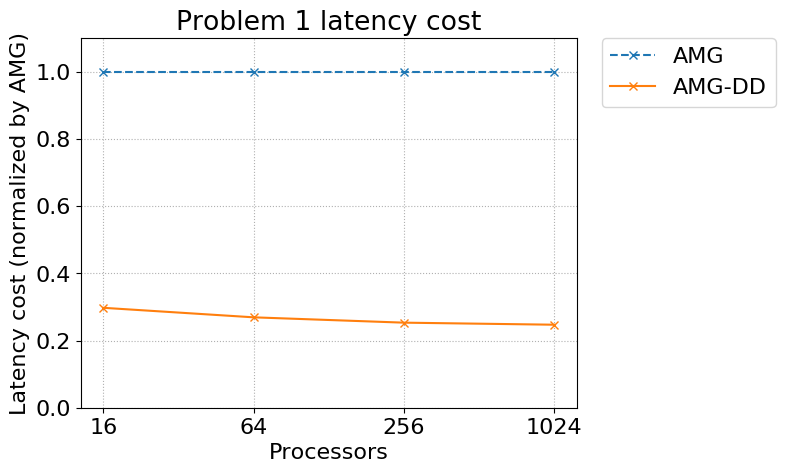}
\includegraphics[height=0.3\textwidth]{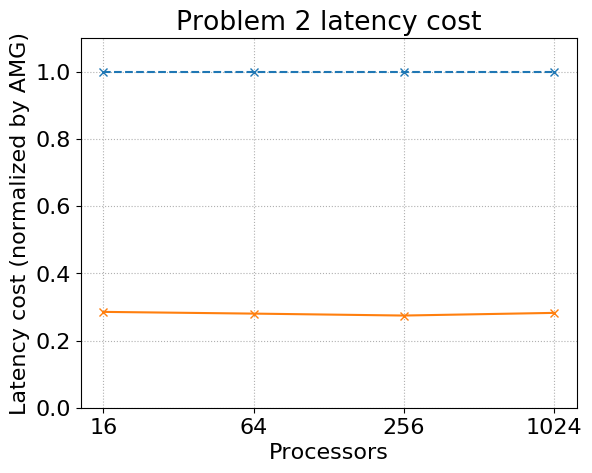}

\includegraphics[height=0.3\textwidth]{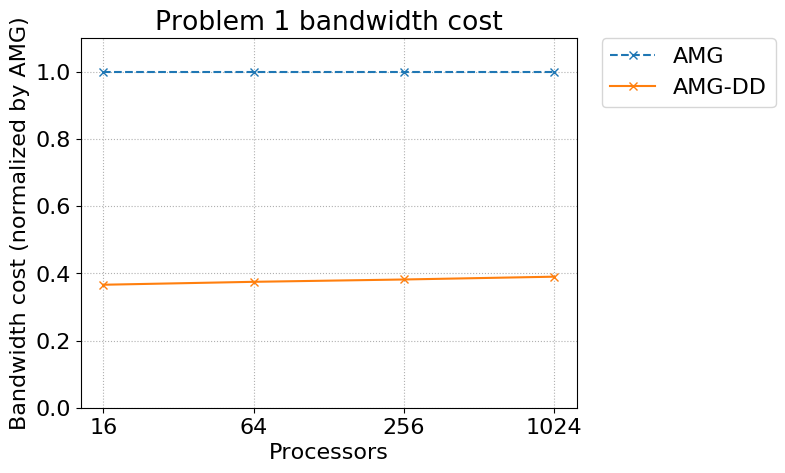}
\includegraphics[height=0.3\textwidth]{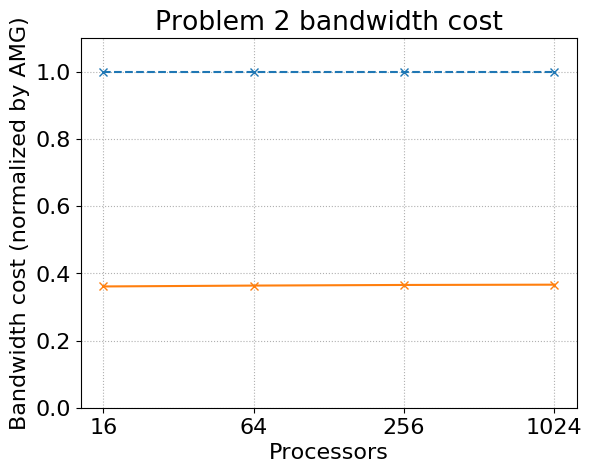}
\caption{Weak scaling of latency costs (top) and bandwidth costs (bottom) for AMG-DD scaled by the costs for AMG for 3D Poisson (left) and rotated anisotropic diffusion (right).}
\label{commWeakScaling}
\end{figure}

Figure \ref{commWeakScaling} shows the communication reduction achieved by AMG-DD over AMG as processor count grows. Large reductions in latency costs and total volume communicated remain very consistent across processor counts. 
Recall that the latency costs for AMG-DD may be directly related to those for AMG V-cycles in the padding 1 case since the communication stencils used for the AMG-DD residual communication algorithm are the same as those used for mat-vecs with $A_l$ on each level. AMG-DD performs a single communication per iteration with these stencils on each level (plus and additional mat-vec on the fine grid and restriction to all levels) whereas AMG performs five communications with the same stencils (plus interpolation and restriction on all levels).
Similar rationale explains the consistent reduction in total communication volume.
The volume communicated by AMG-DD is roughly equal to the number of extra (non-owned) composite grid points on each processor (i.e. the composite-grid overhead). This composite-grid overhead is related to the size of the halo data for $A_l$ on each level, which again, AMG must communicate for each mat-vec, i.e. five times per level.

\begin{figure}
\centering
\includegraphics[height=0.3\textwidth]{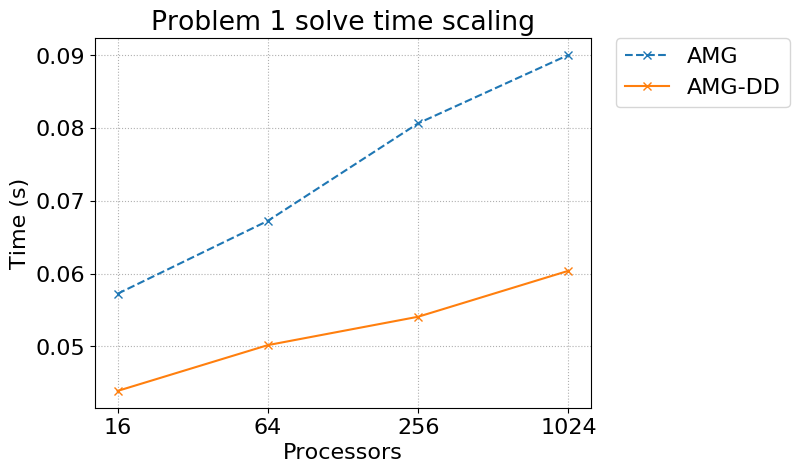}
\includegraphics[height=0.3\textwidth]{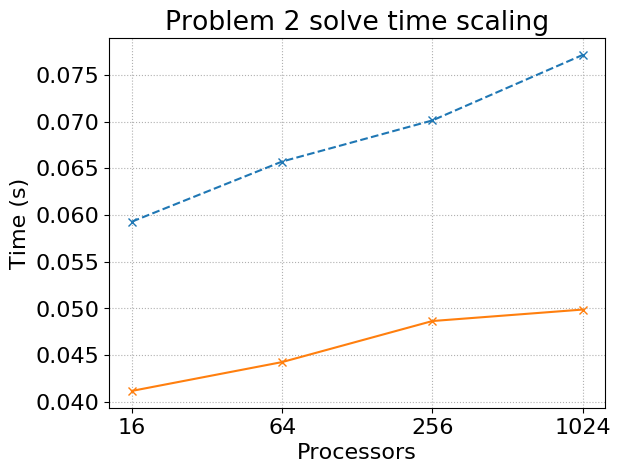}

\caption{AMG and AMG-DD solve time weak scaling for 3D Poisson (left) and rotated anisotropic diffusion (right).}
\label{solveTime}
\end{figure}

\begin{figure}
\centering
\includegraphics[height=0.28\textwidth]{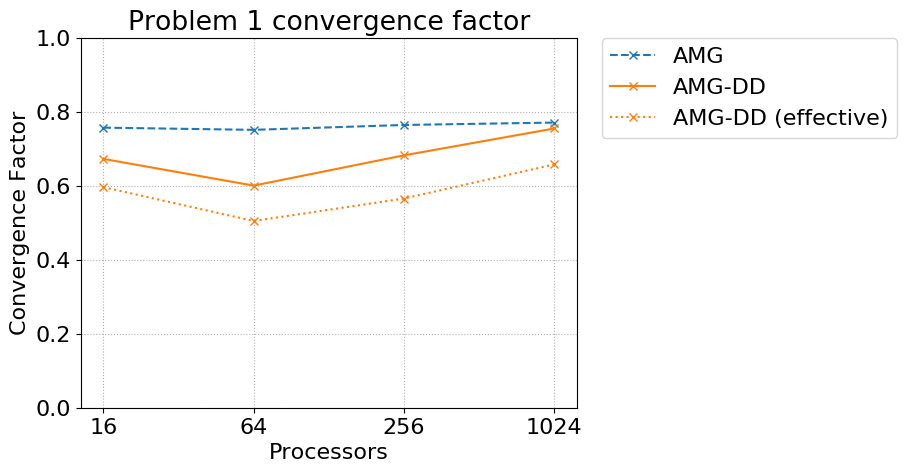}
\includegraphics[height=0.28\textwidth]{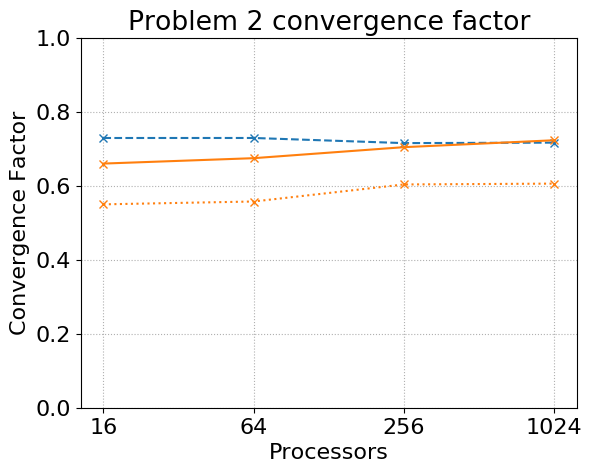}

\caption{Convergence factor weak scaling for AMG vs. AMG-DD (also showing effective convergence factor by time for AMG-DD) for 3D Poisson (left) and rotated anisotropic diffusion (right).}
\label{convFact}
\end{figure}

\begin{figure}
\centering
\includegraphics[height=0.30\textwidth]{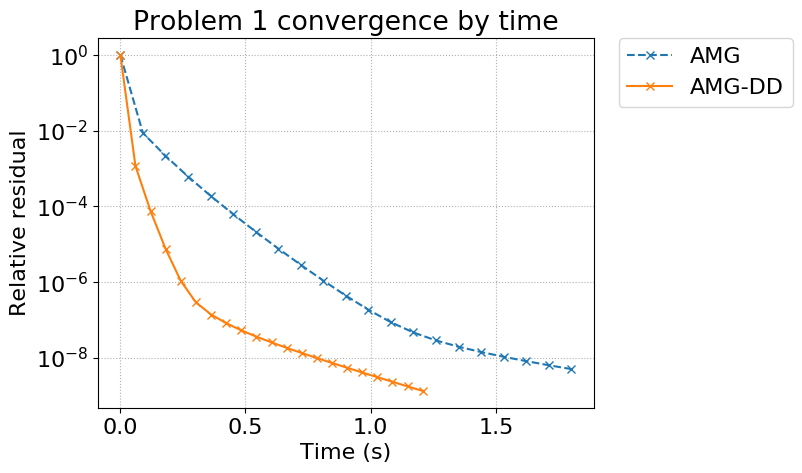}
\includegraphics[height=0.30\textwidth]{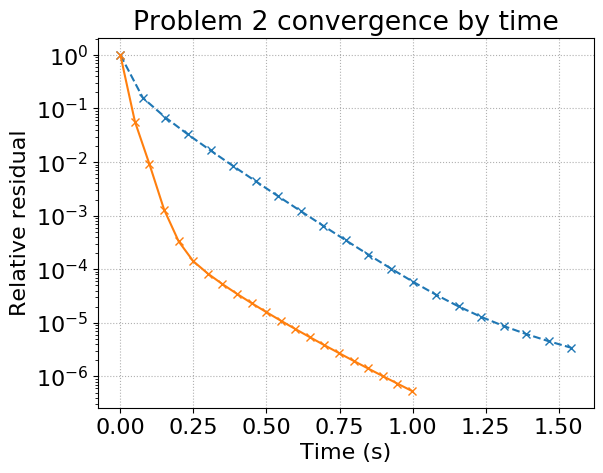}

\caption{Convergence by time for AMG vs. AMG-DD on 1,024 processors for 3D Poisson (left) and rotated anisotropic diffusion (right).}
\label{convByTime}
\end{figure}

Finally, Figure \ref{solveTime} shows solve time weak scaling results for AMG-DD vs. AMG. The timings reported are the minimums achieved over several runs for a single AMG-DD iteration with two inner AlgFAC cycles and for a single AMG V(1,1)-cycle. The methods exhibit similar scaling, with AMG-DD providing significant speedup up to about 1.5x over AMG. Note that AMG-DD achieves this speedup despite doing significantly more computational work compared with AMG, indicating much better utilization of the computational capabilities of the GPU. In addition, Figure \ref{convFact} shows that AMG-DD also achieves better asymptotic convergence factors than AMG for these problems. An "effective" convergence factor for AMG-DD is also shown that incorporates the speedup achieved by AMG-DD over AMG: if $\rho$ is the AMG-DD convergence factor and $S$ is the speedup of AMG-DD over AMG, then the effective convergence factor is $\rho^S$. Thus, the effective convergence factor gives a better illustration of AMG-DD's improvement in terms of convergence per time. Additionally, Figure \ref{convByTime} shows the reduction in relative residual over time in the 1,024 processor case for AMG-DD vs. AMG, most clearly illustrating the advantage of AMG-DD over AMG in terms of accuracy per wallclock time.

\subsection{A more difficult test problem}

The final results of this section are for a more difficult variant of the model problem,
\begin{align}
-\nabla \cdot K \nabla u + du &= f\,, \hspace{1 cm} \Omega \\
u &= 0\,, \hspace{1 cm} \partial \Omega.
\end{align}
Similar to the anisotropic problem above, define $K = Q^TDQ$ and
\begin{align}
Q &= \begin{bmatrix}
\cos(\theta) & \sin(\theta) \\
-\sin(\theta) & \cos(\theta)
\end{bmatrix},
&
D &= \begin{bmatrix}
1 & 0\\
0 & \epsilon
\end{bmatrix}.
\end{align}
For this problem, the domain is a 2D star shape centered on the origin, and different difficulties are included in different parts of the domain by setting
\begin{align}
\epsilon &= 1, \,\,d = 10,000 &x \leq 0, \,\,y \leq 0, \\
\epsilon &= 1, \,\,d = 0 &x > 0, \,\,y \leq 0, \\
\epsilon &= 0.001, \,\,\theta = 0, \,\,d = 0 &x \leq 0, \,\,y > 0, \\
\epsilon &= 0.001, \,\,\theta = 3\pi/16, \,\,d = 0 &x > 0, \,\,y > 0. \\
\end{align}
Thus the lower left quadrant has a strong reaction term, the upper left is simple isotropic diffusion, and the right half plane is anisotropic diffusion with different directions of anisotropy in the upper and lower quadrants. Thus, this test problem (referred to as Problem 3 below) has multiple difficulties localized to different parts of the domain. Again, a random initial guess and zero right-hand side are used. The odd shape of the domain (and subsequently of its partitioning) as well as the different characteristics of the PDE in different parts of the domain are meant to challenge the ability of AMG-DD to represent the global problem on each processor's composite grid.

\begin{figure}
\centering
\includegraphics[height=0.28\textwidth]{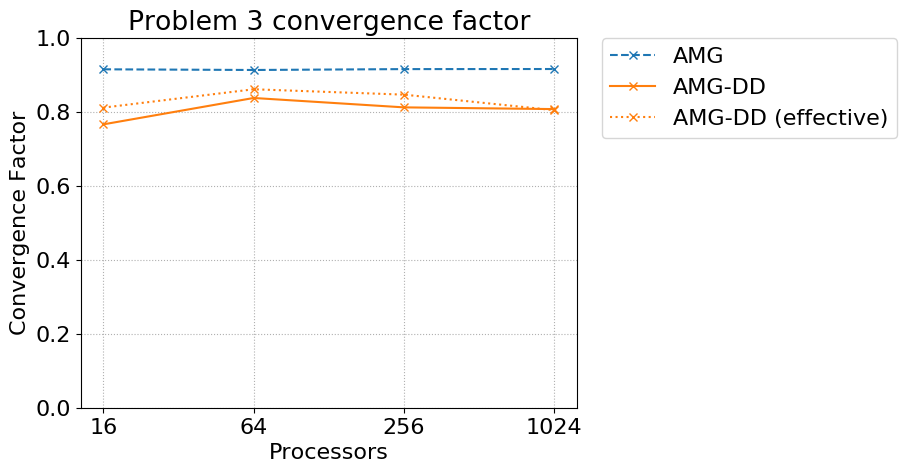}
\includegraphics[height=0.28\textwidth]{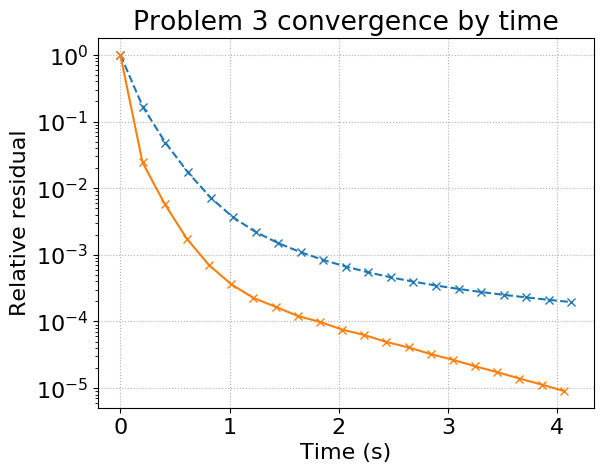}

\caption{Asymptotic convergence factors for different processor counts (left) and convergence by time with 1,024 processors (right) for AMG vs. AMG-DD on Problem 3 with multiple difficulties.}
\label{prob3conv}
\end{figure}

For this problem, there are about 328,000 fine-grid degrees of freedom per processor, and AMG-DD uses 4 inner AlgFAC cycles per iteration. The composite-grid overhead and communication reduction achieved by AMG-DD are very similar to that achieved for the 2D test problem in the previous subsections, so these plots are omitted. Convergence rates for both AMG and AMG-DD are much slower on this more difficult problem, as shown in Figure \ref{prob3conv}. The use of more AlgFAC cycles per AMG-DD iteration for this test case results in AMG-DD iterations that take roughly the same  time as the corresponding AMG V-cycles, but the extra computational work performed by AMG-DD results in swifter convergence. Note that AMG-DD's asymptotic convergence factor of about 0.8 is more than twice as fast as the convergence factor of more than 0.9 for AMG V-cycles. This shows the strengths of AMG-DD in a different light. Although AMG-DD is no faster than AMG in terms of time per iteration in this case, AMG-DD provides an efficient algorithm for performing lots of local computation leading to swifter convergence and ultimately better time to solution compared to AMG.

\section{Conclusions}
\label{conclusions_sec}

This paper continues the development of algebraic multigrid domain decomposition (AMG-DD), a novel, low-communication algorithm built on top of an algebraic multigrid (AMG) hierarchy and designed to achieve similar or better convergence with reduced communication cost compared to AMG V-cycles. An algebraic variant of fast adaptive composite (AlgFAC) cycling is developed that allows for efficient local solution of the composite problems with minimal storage and communication requirements due to the reformulated AlgFAC variant of restriction. Numerical results are shown for the first full parallel implementation of the algorithm, which outperforms AMG V-cycles on a large GPU cluster. Optimal parameters for AMG-DD are discovered via empirical study and then used to produce weak scaling studies that suggest that AMG-DD retains manageable computational overheads, significant communication cost reductions, and superior convergence per iteration compared to AMG as processor counts grow. As a result, AMG-DD is able to achieve better convergence in less time compared with AMG V-cycles on the corresponding hierarchy. In addition, the increased arithmetic intensity of AMG-DD means that it is able to better utilize the local computational power provided by GPU accelerated systems and is thus more suitable for such architectures than traditional AMG.

The underlying AMG hierarchies and smoothers used throughout this paper have been developed and tuned throughout decades of research focused on optimizing AMG V-cycles performance. The next phase of research for AMG-DD will involve developing new strategies for constructing the underlying AMG hierarchy (as well as different methods for AMG-DD composite grid construction) that are best suited to AMG-DD performance. In particular, AMG-DD presents a favorable algorithmic environment for the development of strong smoothers such as block relaxation methods that are difficult to implement or are severely communication-limited in parallel (since smoothing in AMG-DD is performed independently on each processor). Development of such smoothers and their associated coarsening schemes with AMG-DD as the target use-case will be a major area of future research.

\section{Acknowledgments}
The authors acknowledge support by the state of Baden-W\"urttemberg through bwHPC and the German Research Foundation (DFG) through grant INST 35/1134-1 FUGG as well as PRACE for awarding access to Piz Daint at ETH Zurich/CSCS, Switzerland as part of PRACE Preparatory Access Type B from December 2019 to June 2020.
This work was performed under the auspices of the U.S. Department of Energy by Lawrence Livermore National Laboratory under Contract DE-AC52-07NA27344. This document was prepared as an account of work sponsored by an agency of the United States government. Neither the United States government nor Lawrence Livermore National Security, LLC, nor any of their employees makes any warranty, expressed or implied, or assumes any legal liability or responsibility for the accuracy, completeness, or usefulness of any information, apparatus, product, or process disclosed, or represents that its use would not infringe privately owned rights. Reference herein to any specific commercial product, process, or service by trade name, trademark, manufacturer, or otherwise does not necessarily constitute or imply its endorsement, recommendation, or favoring by the United States government or Lawrence Livermore National Security, LLC. The views and opinions of authors expressed herein do not necessarily state or reflect those of the United States government or Lawrence Livermore National Security, LLC, and shall not be used for advertising or product endorsement purposes.

\bibliography{references}

\begin{thebibliography}{10}

\bibitem{Brandt:1985um}
Brandt A, McCormick SF, and Ruge JW.
\newblock {Algebraic multigrid (AMG) for sparse matrix equations}.
\newblock In: Sparsity and its applications. Cambridge Univ. Press, Cambridge;
  1984. p. 257--284.

\bibitem{Stuben:1986vg}
St{\"u}ben K, and Ruge JW.
\newblock {Algebraic Multigrid}.
\newblock In: McCormick SF, editor. Multigrid Methods, vol. 3, Frontiers in
  Applied Mathematics. Philadelphia: SIAM; 1987. p. 73--130.

\bibitem{McCormick:2016vp}
Briggs WF, Henson VE, and McCormick SF.
\newblock {A Multigrid Tutorial, 2nd ed.}
\newblock Edition. SIAM; 2000.

\bibitem{Baker:2012ko}
Baker AH, Falgout RD, Kolev TV, and Yang UM.
\newblock {Scaling hypre's multigrid solvers to 100,000 cores}.
\newblock In: Berry M, editor. High Performance Scientific Computing Algorithms
  and Applications. Springer; 2012. .

\bibitem{Gahvari:2011dh}
Gahvari H, Baker AH, Schulz M, Yang UM, Jordan KE, and Gropp W.
\newblock {Modeling the performance of an algebraic multigrid cycle on HPC
  platforms}.
\newblock In: Proceedings of the 25th International Conference on
  Supercomputing (ICS 2011). Tuscon, AZ; 2011. p. 172--181.

\bibitem{Bienz:2016ii}
Bienz A, Falgout RD, Gropp W, Olson LN, and Schroder JB.
\newblock {Reducing parallel communication in algebraic multigrid through
  sparsification}.
\newblock SIAM Journal on Scientific Computing. 2016;{\bf 38}(5):S332--S357.

\bibitem{Anonymous:KA9t5Fed}
Bank RE, Falgout RD, Jones TM, Manteuffel TA, McCormick SF, and Ruge JW.
\newblock {Algebraic multigrid domain and range decomposition (AMG-DD/AMG-RD)}.
\newblock SIAM Journal on Scientific Computing. 2015;{\bf 37}(5):S113--S136.

\bibitem{Anonymous:ycC62NK_}
Baker AH, Falgout RD, and Yang UM.
\newblock {An assumed partition algorithm for determining processor
  inter-communication}.
\newblock Parallel Computing. 2006;{\bf 32}(5-6):394--414.

\bibitem{Bienz:uc}
Bienz A, Gropp WD, and Olson LN.
\newblock {Node aware sparse matrix-vector multiplication}.
\newblock Journal of Parallel and Distributed Computing. 2019;.

\bibitem{DeSterck:2008fc}
De~Sterck H, Falgout RD, Nolting JW, and Yang UM.
\newblock {Distance-two interpolation for parallel algebraic multigrid}.
\newblock Numerical Linear Algebra with Applications. 2008 Mar;{\bf
  15}(2-3):115--139.

\bibitem{DeSterck:2006et}
De~Sterck H, Yang UM, and Heys JJ.
\newblock {Reducing Complexity in Parallel Algebraic Multigrid
  Preconditioners}.
\newblock SIAM Journal on Matrix Analysis and Applications. 2006 Jan;{\bf
  27}(4):1019--1039.

\bibitem{Yang:2010ii}
Yang UM.
\newblock {On long-range interpolation operators for aggressive coarsening}.
\newblock Numerical Linear Algebra with Applications. 2010;{\bf
  17}(2-3):453--472.

\bibitem{Falgout:2014fw}
Falgout RD, and Schroder JB.
\newblock {Non-Galerkin coarse grids for algebraic multigrid}.
\newblock SIAM Journal on Scientific Computing. 2014;{\bf 36}(3):C309--C334.

\bibitem{Lee:2004cj}
Lee B, McCormick SF, Philip B, and Quinlan DJ.
\newblock {Asynchronous Fast Adaptive Composite-Grid Methods for Elliptic
  Problems: Theoretical Foundations}.
\newblock SIAM Journal on Numerical Analysis. 2004 Jan;{\bf 42}(1):130--152.

\bibitem{Vassilevski:2014eh}
Vassilevski PS, and Yang UM.
\newblock {Reducing communication in algebraic multigrid using additive
  variants}.
\newblock Numerical Linear Algebra with Applications. 2014;{\bf
  21}(2):275--296.

\bibitem{Bank:1999uq}
Bank RE, and Holst M.
\newblock {A new paradigm for parallel adaptive meshing algorithms}.
\newblock SIAM Journal on Scientific Computing. 1999;.

\bibitem{Mitchell:2004hz}
Mitchell WF.
\newblock {Parallel adaptive multilevel methods with full domain partitions}.
\newblock ANACM Applied Numerical Analysis and Computational Mathematics.
  2004;{\bf 1}(1-2):36--48.

\bibitem{Mitchell:2016vg}
Mitchell WF.
\newblock {A parallel multigrid method using the full domain partition}.
\newblock Electronic Transactions on Numerical Analysis. 1998;{\bf 6}:224--233.

\bibitem{Appelhans:2017}
Appelhans DJ, Manteuffel T, McCormick S, and Ruge J.
\newblock A low-communication, parallel algorithm for solving PDEs based on
  range decomposition.
\newblock Numerical Linear Algebra with Applications. 2017;{\bf 24}(3):e2041.

\bibitem{Mitchell:2018fl}
Mitchell WB, and Manteuffel TA.
\newblock {Advances in implementation, theoretical motivation, and numerical
  results for the nested iteration with range decomposition algorithm}.
\newblock Numerical Linear Algebra with Applications. 2018;{\bf
  25}(3):e2149--26.

\bibitem{Vanek:1996dca}
Van{\v e}k P, Mandel J, and Brezina M.
\newblock {Algebraic multigrid by smoothed aggregation for second and fourth
  order elliptic problems}.
\newblock Computing. 1996 Sep;{\bf 56}(3):179--196.

\bibitem{McCormick:1989vx}
McCormick SF.
\newblock {Multilevel Adaptive Methods for Partial Differential Equations}.
\newblock Cambridge University Press; 1989.

\bibitem{Baker:2011ib}
Baker AH, Falgout RD, Kolev TV, and Yang UM.
\newblock {Multigrid smoothers for ultraparallel computing}.
\newblock SIAM Journal on Scientific Computing. 2011 Jan;{\bf
  33}(5):2864--2887.

\end{thebibliography}
\end{document}